\documentclass[onecolumn,notitlepage,superscriptaddress, nofootinbib,nobibnotes, aps,prd,10pt]{revtex4-2}
\usepackage{amsmath,amssymb, fp}  
\usepackage{xcolor}  
\usepackage{soul} 
\usepackage{booktabs}
\definecolor{linkcolor}{RGB}{7,94,84}  
\usepackage[	colorlinks=true,    	
			pdfstartview=FitV,
			linkcolor= linkcolor,
			citecolor= linkcolor,
			urlcolor= linkcolor,
			linktoc=page,
			hyperindex=true,
			hyperfigures=true,  
			breaklinks=true]
			{hyperref}
\usepackage{dblfloatfix}    
\usepackage{balance} 
\usepackage{lastpage}  
\usepackage{mathtools}
\usepackage{braket}
\usepackage{epsfig,rotating,pifont}
\usepackage{graphicx}
\graphicspath{{./figures/}}	
\usepackage[caption=false]{subfig}
\usepackage{placeins}

\usepackage{ragged2e} 
\usepackage{etoolbox}
\usepackage{changepage}   
\usepackage{pgfplots}
\usetikzlibrary{pgfplots.groupplots}
\pgfplotsset{compat=1.3}
\usepackage{tikz}  
\usetikzlibrary{arrows,shapes,positioning}
\usetikzlibrary{decorations.markings,decorations.pathmorphing,decorations.pathreplacing,decorations.text}
\usetikzlibrary{arrows,calc,patterns,shapes.geometric}
\usepackage{fancybox}
\usepackage{verbatim}
\usepackage{multirow}
\usepackage{titlesec}
\usepackage{slashed}

\newsavebox\CBox

\def\dd{\mathrm{d}}

\newcommand{\R}{{\mathbb R}}

\newcommand{\cB}{{\mathcal B}}

\newcommand{\cE}{{\mathcal E}}

\newcommand{\cH}{{\mathcal H}}

\newcommand{\LL}{{\mathcal L}}
\newcommand{\cM}{{\mathcal M}}
\newcommand{\cN}{{\mathcal N}}

\newcommand{\cR}{{\mathcal R}}
\newcommand{\cS}{{\mathcal S}}

\newcommand{\SL}{\mathrm{SL}}
\newcommand{\SO}{\mathrm{SO}}

\renewcommand{\sl}{{\mathfrak{sl}}}
\renewcommand{\so}{{\mathfrak{so}}}

\def\bb#1{\mathbb{#1}}

\def\cw{\curlywedge}

\def\bb{\bar{b}}

\def\be#1\ee{\begin{align}#1\end{align}}
\newcommand{\bea}{\begin{eqnarray}}
\newcommand{\eea}{\end{eqnarray}}
\def\bg#1\eg{\begin{gather}#1\end{gather}}

\def\bsub#1\esub{\begin{subequations}#1\end{subequations}}

\def\q{\quad}

\def\f{\frac}

\definecolor{Green}{RGB}{147,162,153}
\definecolor{Green2}{RGB}{26,148,49}
\definecolor{BrownL}{RGB}{173,143,103}
\definecolor{Red}{RGB}{210,83,60}
\definecolor{BrownD}{RGB}{114,96,86}
\definecolor{GreyD}{RGB}{76,90,106}
\definecolor{GreyB}{RGB}{128,141,160}
\definecolor{Maroon}{RGB}{121,70,61}
\definecolor{Blue}{RGB}{148,184,210}
\definecolor{Blue2}{RGB}{108,144,170}
\definecolor{Blue3}{RGB}{42, 107, 172}
\definecolor{BB}{RGB}{128,184,220}  

\newsavebox\foobox
\usepackage{pict2e}
\makeatletter
\DeclareRobustCommand{\loplus}{\mathbin{\mathpalette\dog@lsemi{+}}}
\DeclareRobustCommand{\lotimes}{\mathbin{\mathpalette\dog@lsemi{\times}}}
\DeclareRobustCommand{\roplus}{\mathbin{\mathpalette\dog@rsemi{+}}}
\DeclareRobustCommand{\rotimes}{\mathbin{\mathpalette\dog@rsemi{\times}}}

\newcommand{\dog@rsemi}[2]{\dog@semi{#1}{#2}{-90,90}}
\newcommand{\dog@lsemi}[2]{\dog@semi{#1}{#2}{270,90}}
\newcommand{\dog@semi}[3]{%
  \begingroup
  \sbox\z@{$\m@th#1#2$}%
  \setlength{\unitlength}{\dimexpr\ht\z@+\dp\z@\relax}%
  \makebox[\wd\z@]{\raisebox{-\dp\z@}{%
    \begin{picture}(1,1)
    \linethickness{\variable@rule{#1}}
    \roundcap
    \put(0.5,0.5){\makebox(0,0){\raisebox{\dp\z@}{$\m@th#1#2$}}}
    \put(0.5,0.5){\arc[#3]{0.5}}
    \end{picture}%
  }}%
  \endgroup
}
\newcommand{\variable@rule}[1]{%
  \fontdimen8  
  \ifx#1\displaystyle\textfont3\else
    \ifx#1\textstyle\textfont3\else
      \ifx#1\scriptstyle\scriptfont3\else
        \scriptscriptfont3\relax
  \fi\fi\fi
}
\makeatother

\usepackage{etoolbox}

\makeatletter
\patchcmd{\ttlh@hang}{\parindent\z@}{\parindent\z@\leavevmode}{}{}
\patchcmd{\ttlh@hang}{\noindent}{}{}{}
\makeatother

\newcommand{\secmark}{}
\newcommand{\marktotoc}[1]{\renewcommand{\secmark}{#1}}

\makeatletter
\renewcommand{\@dotsep}{1000}

\titleformat*{\section}{\center\bfseries}

\def\lp{\ell_\text{Pl}}

\def\Sym{\Omega}    
\def\Presym{\Theta}  
\def\presym{\theta}  
\newcommand{\ls}{{\ell_{\rm{s}}}}
\def\centr{\mathfrak{c}}
\def\ri{{r_\text{i}}}
\def\rf{{r_\text{f}}}
\def\vi{{v_\text{i}}}
\def\vf{{v_\text{f}}}
\def\bb{B}
\def\xx{X}

\begin{document}

\title{Phase spaces and symmetries of Vaidya superspace}

\author{Salvatore Ribisi}
\affiliation{Aix Marseille Univ, Université de Toulon, CNRS, CPT, Marseille, France}
\author{Francesco Sartini}
\affiliation{Okinawa Institute of Science and Technology Graduate University,
1919-1 Tancha, Onna-son, Okinawa 904-0495, Japan}

\date{\today} 
 
\begin{abstract}
We investigate the classical symmetries of the dynamics of the null-dust spherically symmetric Vaidya spacetime. Einstein's equations for this model can be obtained as equations of motion of a two-dimensional field theory. We discuss the transformations leaving invariant such equations of motion. These are given by two distinct sets, the residual diffeomorphisms coming from general relativity and the generalisation of the Schr\"odinger symmetry, recently found for the static Schwarzschild black holes. Surprisingly, these two sets represent the symmetries of two different action functionals, leading to the same equations of motion, but with different phase spaces.
\end{abstract}

\maketitle  

\makeatletter

\makeatother

General relativity rests strongly on a symmetry principle, the invariance under diffeomorphisms, corresponding to coordinate changes on spacetime. However, these symmetries appear as gauge symmetries, redundancy of the physical description. Hence, they are not associated with any physical content. The Noetherian duality between symmetries and conserved quantities, or charges, would give zero charge for gauge symmetries, leading to the cumbersome task of defining observables in gravity. The situation changes dramatically in the presence of boundaries, may they be asymptotic or at finite distance. They can promote some gauge symmetries to have a non-zero charge living on codimension-two \textit{corners} of spacetime \cite{Regge:1974zd,Carlip:1994gy,Balachandran:1995qa,Freidel:2020xyx,Freidel:2020svx,Freidel:2020ayo,Donnelly:2020xgu,Freidel:2021cjp}

Thinking of boundaries as bridges between different regions makes the corner symmetry algebra very relevant for the study of entanglement entropy between subregions. In this spirit, we can aim to use the representation theory of the corner algebra as non-perturbative handles on quantum gravity \cite{Freidel:2021cjp}; but it also allows us to, more conjecturally, make spacetime and its topology emerge from quantum entanglement between subregions \cite{Carrozza:2021gju,Carrozza:2022xut}.

However, even letting aside the boundaries, bulk symmetries might have their relevance. Some transformations link different sets of bulk solutions in general relativity. For example, the Newman-Janis algorithm allows obtaining rotating black hole solutions, out of the Schwarzschild one, through a complex coordinate transformation \cite{Newman:1965tw,Drake:1998gf}. In addition, we have regularities in the tower of quasi-normal modes or responses to perturbation that come from approximate near-horizon symmetries \cite{Chen:2010ik,Kim:2012mh}. The latter is also related to boundary structure and black hole entropy \cite{Birmingham:1998jt,Carlip:1998wz,Carlip:2017xne}

Recently a very peculiar class of symmetries \cite{Geiller:2020xze,BenAchour:2022fif,Geiller:2022baq,BenAchour:2023dgj}, for some very regular solutions of general relativity, has drawn some attention. These highly-symmetric spacetimes, or \textit{minisuperspaces}, can be described as mechanical models, focusing on the evolution in just one spacetime direction and freezing the other ones. This is analogous to selecting the zero modes of geometry, but despite seeming very simple at first glance, they are relevant for cosmological or near-singularity applications.

These symmetries fully encode the evolution of the physical spacetime, and have an elegant interpretation in terms of  \textit{geometrization} of the dynamical space. The configuration field space is endowed with a metric, constructed out of the kinetic term of the reduced action \cite{Geiller:2022baq,BenAchour:2022fif}. 

Originally discovered for the isotropic cosmological setup \cite{BenAchour:2017qpb,BenAchour:2019ywl}, these minisuperspace symmetries have also been uncovered for black hole models \cite{Geiller:2020xze,BenAchour:2022fif,BenAchour:2023dgj} and anisotropic cosmologies \cite{Geiller:2022baq,Sartini:2022ecp}. A review of a systematic approach to the exploration of homogeneous models can be found in \cite{Geiller:2022baq,Sartini:2022ecp} or see \cite{BenAchour:2022fif} for an equivalent technique, known as Eisenhart--Duval lift, based on an extended phase space \cite {Cariglia:2016oft}.

The interest in minisuperspaces goes beyond the elegant relationship between symmetries, dynamics and geometrical structure of the field space. Recent works have pointed out that astrophysically relevant models possess a symmetry group equivalent to the Schr\"odinger group. This conformal group has a key role in non-relativistic hydrodynamics and for some Bose-Einstein condensates. It suggests an intriguing correspondence between the response to perturbation of these gravitational systems and fluid analogues.
 
Moreover, there seems to be an intriguing feature of the minisuperspace symmetries in relationship with the boundaries of spacetime. In all these models we need indeed a regulator to deal with an infinite homogeneous slice. This turns out to interplay with the symmetries, being modified by them \cite{Geiller:2020xze,Sartini:2022ecp}. However, to better understand these structures we should go beyond the simple homogeneous setup, by including inhomogeneity in cosmologies or non-stationary processes for black holes.

In this paper, we will discuss the extension of the Schr\"odinger symmetry to the simplest non-static generalisation of the Schwarzschild solution, known as Vaidya spacetime. This will force at least one field to evolve in two directions. On top of the radial dependence of the system, already considered for the stationary minisuperspace, we will add the dependence on a null coordinate. The richer spacetime structure makes the residual diffeomorphism gauge freedom less trivial than the static case. We will then discuss how this gauge symmetry interplays with the Schr\"odinger transformations.

The paper is organised as follows. We started in section \ref{sec:Vaidya action} by introducing the Vaidya superspace as a two-dimensional field theory, coming from the spherical symmetric general relativity in a particular gauge. After verifying that the solutions of the equations of motion are consistently given by the Vaidya solutions alone, we will move to the discussion about their symmetries. In section \ref{sec:symm} we introduce these symmetries simply as the transformations mapping solutions into solutions. However, to interpret them in a Noetherian sense, we shall give a notion of phase space. In \ref{sec:sympl} we show that the two sets of transformations, Schr\"odinger and gauge symmetries, are integrable on two different phase spaces, coming from two different theories, leading to the same classical equations.

Indeed the plural \textit{phase spaces}  in the title of this article is not a typo, we can obtain the same classical spacetime from two inequivalent phase spaces coming from different action functionals.



\section{Action and equations of motion of Vaidya superspace}
\label{sec:Vaidya action}
The Vaidya metric represents the simplest radiating solution for black holes and provides a natural testbed to address questions related to black hole evaporation \cite{Bicak:1997bx,Louko:1997wc,Campiglia:2016fzp,Hajicek:2002ny,Eyheralde:2017jzd}. The presence of hydrodynamical symmetries in such a model might give new insights into the problem. 

The metric is usually presented in the Eddington--Finkelstein gauge, generalising the spherically symmetric ansatz by breaking the stationarity or equivalently partially breaking the diffeomorphism invariance on the radial-temporal plane. More in detail, we take a spherically symmetric ansatz in four spacetime dimensions, which means separating the angular directions with respect to the other two coordinates. The latter represents the radial-temporal plane, whose compactification is the Penrose diagram. On top of this, we force one of the coordinates to be null, imposing the gauge condition $g_{rr}=0$. The ansatz that we take is thus
\be
    \dd s^2 =\f{\bb(v,r)}{\xx(r,v)} \dd v^2 + 2 N(r,v) \dd v \dd r + \xx(r,v)^2 \dd S_{(2)}^2\,,
    \label{metric}
\ee
where the term $\dd S_{(2)}^2$ represents the usual two-sphere metric. The choice of parametrization for the $g_{vv}$ term is chosen in this way to simplify the notation below. We insist again on the fact that this ansatz partially breaks the covariance along the mixed $r$-$v$ direction, because it does not contain the $rr$ term, and constrains $v$ to be a null coordinate. In other words, the diffeomorphisms that preserve the ansatz \eqref{metric} are the ones generated by the vector fields
\be
\xi = \xi^r(r,v)\partial_r+\xi^v(v)\partial_v + \sigma[S_{(2)}] \label{diffeo}\,,
\ee
with $\sigma$ generating the celestial sphere's global $\SO(3)$ rotations. It has a non-null component only along the angular direction, trivially commuting with the null and radial diffeomorphisms. We will later show that the partial breaking of the full two-dimensional diffeomorphism group on the $r$-$v$ plane will have the consequence of losing one Einstein equation, corresponding to the mass conservation. On the other hand, this allows us to obtain the non-static Vaidya metric as a solution. We shall remark that in principle our ansatz contains both the emitting and absorbing Vaidya pure radiation fields\footnote{With the notation chosen here we have an absorbing field for positive $N$, for which we usually use the ingoing null coordinate $v$. On the contrary, the outgoing null coordinate is typically denoted by the letter $u$. We choose here to keep $v$ for the null coordinate, also in the ingoing case. We use the sign of the field $N$ to flip between absorbing and emitting cases}, depending on the sign of $N$. 

We would like to obtain Einstein's equation from the variational principle of some action functional. The most obvious being the reduced Einstein--Hilbert action 
\be
    \cS_\text{EH} &= \f{1}{16 \pi \lp^2}\int \dd^4 x \sqrt{|g|} \cR \notag \\ 
    &= \frac{1}{2\lp^2} \int \dd v \dd r \left[ N + \frac{\xx' (2\partial_v(N\xx)-\bb\xx')}{N} \right] + \left.\f{1}{4\lp^2} \int \dd v \left[ \f{\partial_r (\bb \xx^3)-2\xx^2 \dot N}{N \xx^2}\right]\right|_{\ri}^{\rf}- \left.\f{\xx^2 }{2 \lp^2} \right|_{\ri,\vi}^{\rf,\vf}\,,
\label{action_bg}
\ee
where the dot represents the derivative with respect to $v$ and the prime with respect to the radial direction. The four-dimensional Ricci scalar is denoted with $\cR$, and we choose units such that the four-dimensional Newton constant is $G=\lp^2$, with the Plank length $\lp$. The action must be thought to describe the variational problem in the region between two slices at a constant radius and two null surfaces at constant $v$. However, for the moment let us neglect the discussion about boundary conditions and just focus on the bulk equation of motion.

The first part of the action \eqref{action_bg} contains the bulk Lagrangian and the second term is a boundary term, acting on the polarization of the phase space (when evaluating the evolution in the $r$ direction). The last one is a corner term for the symplectic potential and might play a role in the determination of the charges and algebra. However, for the study of the classical solutions and bulk symmetries, the only relevant part is given by the bulk term.  We can thus evaluate the variational principle of the two-dimensional field theory action 
\be
\cS_0 =\frac{1}{2\lp^2} \int \dd v \dd r \left[ N - \frac{\xx' (\bb'-2\partial_v(N\xx))}{N} \right]\label{action_0} .
\ee

The first property that we shall verify is the consistency with general relativity. The fact that the Euler--Lagrange equations for the Lagrangian \eqref{action_0} give Einstein's equations for the metric \eqref{metric} is a highly non-trivial statement. Indeed we will show that one equation is missing and the mass will be allowed to evolve in the null direction. A straightforward computation gives us the equations of motion for the field theory \eqref{action_0}:
\bsub\be
    0&\approx \partial_r \left(\xx'/N\right)  \,,\\
    0&\approx N(\bb\xx''-4 N \dot \xx')-\bb\xx'N'-2 \xx(N \dot N'-\dot N N')\,, \\
    0&\approx N^2 + \xx'\bb\xx'-N\partial_v\partial_r \xx^2 \,. \label{scalar_constr}
\ee \label{eoms} \esub
These are not all independent, because of the residual gauge freedom generated by \eqref{diffeo}. We can analytically solve the equations of motion for any function $\xx$, that will be later identified as a dilatonic field from a two-dimensional perspective \cite{Grumiller:2002nm,Afshar:2019axx,Ruzziconi:2020wrb,Grumiller:2021cwg}. We can deparametrize the evolution with respect to this field, and add two free functions depending only on $v$ as initial conditions. The general solutions of \eqref{eoms} are\footnote{We use $\approx$ to denote on-shell equalities}
\be
    \bb \approx B_0(v)- \xx\,n(v)^2+ 2n(v) \xx \dot \xx\,,\q \q N\approx n(v) \xx' \,.
\label{solutions} \ee 
As expected, the solution space is not completely invariant under spacetime reparametrization, but only under the action of the residual diffeomorphisms \eqref{diffeo}. Among them, only the radial and null directions act non trivially on the solution space, while the celestial sphere angular directions are gauged out of the model. The coordinates $r$ and $v$ play two different roles, and our ansatz \eqref{metric} is invariant only under diffeomorphisms that leave $v$ as a null coordinate. As already announced, from the variation of the action we miss the mass conservation, leading to an on-shell Vaidya metric. Replacing the solutions \eqref{solutions} into the ansatz \eqref{metric} we explicitly get
\be
    \dd s^2 &= - \left(n^2- \f{B_0}{\xx}-2n \dot \xx\right) \dd v^2 + 2 n \xx' \,\dd v \dd r + \xx(r,v)^2 \dd S_{(2)}^2\notag \\
    &= - \left(1- \f{2G M(v)}{\xx}\right) n^2 \dd v^2 + 2 n \dd v\, \dd \xx + \xx^2\, \dd S_{(2)}^2\,,\q\q 2\lp^2 M(v) =B_0/n^2\,. \label{on_shell_metric}
\ee
The usual null coordinate of Eddington--Finkelstein parametrization of Vaidya is given by $\pm \dd V = n(v) \dd v$, while $\xx$ represents the radial coordinate \footnote{The variational principle of the action \eqref{action_0} allows for both positive and negative $N$, representing emitting and absorbing Vaidya spacetimes. Moreover, the residual diffeomorphisms allow to flip the sign of $N$ by exponentiation of a $\xi$ pointing backwards in the $v$ direction}. 

The missing equation, which should come from the variation of the action with respect to the $g^{vv}$ term\footnote{The vanishing of the $g_{rr}$ term is, of course, equivalent to the vanishing of $g^{vv}$}, is the mass conservation. Indeed the on-shell Einstein tensor for our ansatz has a non-vanishing component
\be
G_{vv} \approx \f{n \dot B_0- 2 B_0 \dot n}{\xx^2 n^2}= \f{2\lp^2\dot M}{\xx^2}\,.
\ee
This means that the variational principle of the action \eqref{action_0} gives non-vacuum Einstein's equations for the bulk ansatz \eqref{metric}, effectively coupled to a pure radiation field with stress-energy tensor satisfying 
\be
T_{\mu\nu} \approx \f{1}{4 \pi} \f{\dot M}{X^2} l_\mu l_\nu\,,
\ee 
with the null form $l_\mu \dd x^\mu =-\dd v$. The full diffeomorphism invariance on the $r-v$ plane can be restored by adding such null dust to the Lagrangian \cite{Bicak:1997bx,Louko:1997wc}. This will give the Vaidya solution for an ansatz in which $v$ is not necessarily null, adding a $g_{rr}$ term, whose variation will impose the total mass conservation. We chose here to hide the matter contribution in the partial gauge fixing, to make the comparison with the static case easier to handle.

\subsection{Relationship with two-dimensional dilatonic gravity}
\label{sec:2d_dilaton}
Before moving to the discussion about the classical symmetries of this model, let us open a small parenthesis on an interesting relationship between this model and a general two-dimensional dilatonic theory. A general class of such theories, whose dynamical content is given by a two-dimensional metric $g_{\mu\nu}^{(2)}$ and a scalar field $\Phi$, is given by \cite{Ruzziconi:2020wrb,Grumiller:2021cwg}
\be
 \cS_\text{DGT}=\f{1}{16 \pi G_{2D}} \int \dd^2 x \sqrt{|g^{(2)}|}\left(\Phi R -U(\Phi) (\nabla \Phi)^2 -2V(\Phi) \right)\,,\label{dgt}
\ee
where $U$ and $V$ are functions of the dilaton field $\Phi$, and $R$ is the Ricci scalar of the two-dimensional metric. Within this class, we have models such as JT gravity ($U=0$, $V= \Lambda \Phi$) or the CGHS model ($U=0$, $V=\lambda$). The four-dimensional spherically symmetric gravity also belongs to this class. Indeed we can identify the two-dimensional metric as the one describing the radial and temporal direction (i.e. the Penrose diagram), while the scalar field determines the measure of the celestial sphere (points on the Penrose diagram). Let us consider a general four-dimensional spherically symmetric ansatz 
\be
\dd s^2_{4D} = e^{2\Omega(\Phi)} g^{(2)}_{\mu\nu}\, \dd x^\mu \dd x^\nu  + \Phi\, \dd S \,,\q \mu,\nu\in\{0,1\}\,. \label{4Dss}
\ee

The two-dimensional metric can be identified up to a conformal factor $\Omega$, taken to be a function of the scalar dilaton. The conformal rescaling interplays with the potential $U$, $V$ and we can use this fact to set one of the potentials to zero. The Einstein--Hilbert action of the spherically symmetric line-element \eqref{4Dss} is 
\be
 S_\text{4D} &= \f{1}{16 \pi \lp^2}\int \dd^4 x \sqrt{|g^{(4D)}|} \cR \notag \\ 
    &=  \f{1}{4 \lp^2}\int \dd^2 x \sqrt{|g^{(2)}|} \left( \Phi R +2e^{2\Omega} +\f{1}{2 \Phi} (\nabla \Phi)^2-2 \Phi \nabla^2 \Omega \right) +\text{“boundary terms"}\,.
\ee 
We can eliminate the kinetic term for the dilaton (up to a boundary term)  by choosing
\be
e^{2\Omega}=\f{1}{\sqrt \Phi}\,.
\ee
This identifies the reduced four-dimensional action as the one in the class \eqref{dgt} with $U=0$, $V=-1/\sqrt \Phi$. At this point, this model still possesses the whole invariance under two-dimensional diffeomorphism, but we can immediately recognise that our ansatz \eqref{metric} corresponds to the Bondi gauge of the two-dimensional metric, and the field $\xx$ is the square root of the dilaton. Taking the two-dimensional metric $g_{\mu\nu}^{(2)}$ in the Bondi gauge
\be
\dd s_{2D}^2=  B(v,r) \dd v^2 + 2  \xx(r,v) N(r,v) \dd v \dd r\,\label{metric2}
\ee
and plugging this ansatz in the dilatonic action, with the choices  $U=0$, $V=-1/\xx$, $\Phi=\xx^2$, gives the “Vaidya action" \eqref{action_bg}, up to a boundary term, meaning that
\be
S_\text{EH} = \f{1}{4 \lp^2} \int \dd^2 x \sqrt{|g_{2D}|}\left(\xx^2 R^{(2D)} +\f{2}{\xx} \right) + \f{1}{4\lp^2} \int \dd v \f{B \xx'}{N}- \left.\f{3 \xx^2 }{4 \lp^2} \right|_{\ri,\vi}^{\rf,\vf}\,.
\ee
As a consequence, this observation implies that one must be extremely careful when plugging the Bondi gauge in off-shell quantities. As shown here, if we do so in the two-dimensional case at the action level and \textit{then} we evaluate the Euler--Lagrange equation for the gauge fixed action, we will lose the mass conservation equation. Although here we want precisely to use this fact to obtain the Vaidya solution from the “wrong" gauge fixing without specifying the matter content of the theory, this has important consequences for the study of asymptotic symmetries.

\subsection{Linear radial gauge}
\label{sec:2d_dilaton}
Inspired by the previous works on Schwartzschild black holes \cite{BenAchour:2022fif,Geiller:2020xze,Sartini:2022ecp}, we do a further gauge choice by picking a linear square root of the dilaton, i.e. imposing 
\be
\xx(r,v)= A_0(v) \left(r-\phi_0(v)\right)\,.
\ee 
By doing so, we obtain a solution space spanned by four free functions depending on the null coordinate. The same happens in \cite{Ruzziconi:2020wrb}, where they study the integrability of asymptotic large diffeomorphisms in two-dimensional gravity, even if therein the linearity condition is on $\Phi=\xx^2$. The choice here is taken as a natural generalisation for the previous works in black hole minisuperspaces and it is implemented by the condition $N'=0$, which is the same as the one chosen in \cite{Geiller:2020xze,BenAchour:2022fif}, while the one in \cite{Ruzziconi:2020wrb} is implemented by $\partial_r(X N)=0$.

The equations of motion get rewritten in this gauge as
\bsub\be
    0&\approx\xx'' \label{eom_X}  \,,\\
    0&\approx \bb''-4 N \dot \xx'\,, \label{eom_B}\\
    0&\approx N^2 + \xx'\bb'-N\partial_v\partial_r \xx^2  \label{eom_N}\,,
\ee \label{eoms_lin} \esub
and the solution space is now spanned by four functions of the null coordinate alone $A_0$, $B_0$, $\phi_0$ and $n$,
\be
X&\approx A_0(v) \left(r-\phi_0(v)\right)\,,\notag \\
B&\approx B_0(v)-A_0\, n \left(r-\phi_0\right)(n -2 (r-\phi_0) \dot A_0 +2 A_0 \dot \phi_0)\,, \\
N&\approx A_0(v)\, n(v)\,.\notag 
\ee

\section{Moebius symmetry and residual diffeomorphisms}
\label{sec:symm}
With the solution space at hand, we can now turn to the study of the symmetries of this model. In the first place, we will search for sets of transformations that preserve the equations of motion, in the sense that they map solutions of \eqref{eoms_lin} into solutions. We also work within the linear gauge $N'=0$. 

This section aims to show how it is possible to generalise the minisuperspace Schr\"odinger symmetry, originally found for the static black holes to the Vaidya model. For this let us recall that the two-dimensional Schr\"odinger group splits into the semi-direct product  
\be
\rm{Sh}(2) = \left (\SL(2,\R) \times \SO(2) \right ) \ltimes \left (\R^2 \times \R^2\right )\,,
\ee
where the algebra generating the abelian part $(\R^2 \times \R^2)$ contains a central extension. 
In quantum mechanics, where the group has been introduced in the first place, the central charge is the Plank constant. Classically, the abelian symmetry group corresponds to constant shifts of position and momenta of a free particle. The $\SO(2)$ part represents the rotations of the two-dimensional plane, and $\SL_2$ generates conformal symmetries of the Schr\"odinger equation.

In particular, the realization of the $\SL(2,\R) \times \SO(2) $ subgroup on the black hole superspace translates into a Moebius transformation on the radial coordinate, while the metric coefficients transform as conformal fields of different weights. At the level of spacetime, this corresponds to an anisotropic Weyl rescaling of the geometry and is not a residual diffeomorphism \cite{Geiller:2022baq,Sartini:2022ecp}.  To begin, we will focus on this subgroup to try to extend the symmetry to the Vaidya superspace.

In the minsuperspace setup, the Schr\"odinger symmetry emerges naturally from a second geometrization procedure, mapping the spacetime dynamics to a point particle geodesic motion on the field space. From this perspective, the symmetries are associated with conformal properties of the supermetric \cite{BenAchour:2022fif,Geiller:2022baq}. In the case we study here, we lack such a point particle interpretation, because of the presence of an infinite dimensional field space. However, we can rewrite the equations of motion in a similar way to the finite-dimensional space. We can achieve this by redefining the fields to get rid of the null direction derivatives in the equations of motion. While this is already the case for $\xx$ in \eqref{eom_X}, it is convenient to introduce the quantity
\be 
\cB := \bb - 2r N\dot X \,.
\ee
It's easy to convince ourselves that the equation of motion \eqref{eom_B} for $\bb$ is rewritten in a very simple way for the new field, namely
$\cB '' \approx 0$. Although this field redefinition makes the equations of motion have a nicer form, its geometrical interpretation is not straightforward.

It also turns out to be useful to redefine the field $N$ in a way that makes the last equation more compact. Let us consider the combination
\be
\cN^2 := N^2 -2 N\dot \xx' (\xx - r \xx')\,.
\ee
Combining the gauge condition $N'=0$ and the equation of motion $\xx''=0$, we can show that also $\cN$ is independent of the radial coordinate. Equivalently, on our solution space, $\cN' \approx 0$. This brings us to make the $v$ derivatives completely disappear from the equation of motions, turning \eqref{eoms_lin} into
\bsub\be
    0&\approx\xx''  \label{eom_XX}\,,\\
    0&\approx \cB'' \label{eom_BB}\,,\\
    0&\approx \cN^2 + \xx'\cB'\,.\label{eom_NN}
\ee \label{eoms_lin2} \esub
In particular, the first two equations decouple the evolution of $\xx$ and $\cB$ and are both in a form which is invariant under Moebius reparametrization of the radial coordinate\footnote{The same form indeed appears in the minisuperspace setup, once we write the field space in the appropriate null variables \cite{Geiller:2022baq}, and it is related to the conformal invariance of the free particle mechanics}. Let us define the transformation
\bsub \be
    r \ &\rightarrow \ \tilde{r} = h(r) := \frac{\alpha \ r + \beta}{\gamma \ r + \delta} , \qquad \alpha \delta - \beta \gamma = 1\,,\q h'(r) = \f{1}{(\gamma r +\delta)^2}\,, \\
    \xx(r,v) \ &\rightarrow \ \widetilde{\xx}(\tilde r,v) = \lambda \sqrt{h'(r)}  \ \xx(r,v)\,, \q\q \lambda=const\,,\\
    \cB(r,v) \ &\rightarrow \ \widetilde{\cB}(\tilde r,v) = \lambda^{-1} \sqrt{h'(r)}  \ \cB(r,v)\,, 
\ee \label{transf_moebius}
\esub
This is easily shown to leave the first two equations of motion \eqref{eom_XX} and \eqref{eom_BB} invariant, we have indeed e.g. the first one
\be
\xx'' \mapsto \partial_{\tilde r}^2 \widetilde X= \lambda \f{h'^2 \xx''+ X( 2 h' h''' -3 h''^2)/4}{ \sqrt {h'}}= \lambda h'^{3/2}\, \xx''\,,
\ee
recognising the Schwarzian derivative, that vanishes for the Moebius transformation above,
\be
\text{Sch}[h]:= \f{h'''}{h'}-\f{3}{2}\left(\f{h''}{h'}\right)^2 =0 \q \Leftrightarrow \q h= \frac{\alpha \ r + \beta}{\gamma \ r + \delta}\,.
\ee
The same happens for the equation \eqref{eom_BB} concerning the evolution of $\cB$. 

We now need to discuss the last equation of motion and the transformation for $\cN$. The most naive way of defining how $\cN$ transform, is precisely through the equation of motion \eqref{eom_NN}. Indeed, for the transformations to be symmetries, we must have
\be
\widetilde {\cN} \approx \left(-\partial_{\tilde r} \widetilde \xx \partial_{\tilde r} \widetilde \cB \right)^{1/2} = \f{1}{\sqrt{h'}}\left(-\xx' \cB' -\f{1}{2} \partial_r\left(\f{h''}{h'}\xx \cB \right)\right)^{1/2}\approx 
\f{1}{\sqrt{h'}}\left(\cN^2 -\f{1}{2} \partial_r\left(\f{h''}{h'}\xx \cB \right)\right)^{1/2}\,. \label{new_constraint}
\ee
We can use the last expression to define $\widetilde \cN$. This expression maps $\cN$ to some $r$ dependent field, seeming to break the gauge condition. However, once it is projected onto the solution space, it happens to be consistent with the gauge choice. If the equations \eqref{eoms_lin2} hold, and $\cN'=0$, then we also have $\widetilde \cN'=0$

We can also work with on-shell quantities and find a closed expression for the symmetry flow on the solution space (see also appendix \ref{app:symm_long}). This takes a very compact form in terms of a particular combination of initial conditions. Let us define the quantities
\bsub \be
\psi_0 &:= \frac{n \partial_v (\phi _0 A_0)^2+n^2 \phi _0 A_0  +B_0}{A_0 n (2 \phi _0\dot A_0+n)}\,,\\
Q_0 &:= n A_0 (2 \phi _0\dot  A_0+n)\,,
\ee \esub
corresponding respectively to the zero and (minus) the first derivative of the on-shell field $\cB$. The $\SL(2,\R) \times \SO(2) $ group maps solutions for the fields $\xx$ and $\cB$ into solutions, and it acts non-trivially on the solution space as
\bsub \be
A_0     &\to \f{\lambda A_0}{\sqrt{h'(\phi_0)}}\,,\\
\phi_0  &\to h(\phi_0)\,,\\
Q_0     &\to \f{Q_0}{\lambda \sqrt{h'(\psi_0)}}\,,\\
\psi_0  &\to h(\psi_0)\,,
\ee \label{on_shell_mob}\esub
for the Moebius function $h$ defined in \eqref{transf_moebius}. It is also useful to write the corresponding infinitesimal transformations, both on the field and solution spaces. We evaluate them at the same space-time point, meaning that we define the variation of a field $\chi(r,v)$, or of a solution space parameter $\psi(v)$ as
\bsub \be
\delta\chi &:=\tilde \chi(r,v)-\chi(r,v) = \tilde \chi(\tilde r,\tilde v)-\chi(r,v)-\delta r\, \chi' -\delta v\, \dot \chi \,,\\
\delta\psi &:=\tilde \psi(v)-\psi(v) = \tilde \psi(\tilde v)-\psi(v) -\delta v\, \dot \psi \,.
\ee \esub
Let us remark that, for both fields $\xx$ and $\cB$, we have assumed that the finite conformal transformations \eqref{transf_moebius} leave the null coordinate invariant. We set then $\delta v=0$, and we will discuss later the residual null reparametrization, coming from the reduction of gauge diffeomorphisms \eqref{diffeo}. The infinitesimal generator of the Moebius transformation is given by a second-degree polynomial \cite{Geiller:2020xze}, while $\lambda$ differs from the identity by a small constant,
\be
h(r)&\sim r +\epsilon(r)\,,\q\q \epsilon'''=0\,,\notag\\
\lambda &\sim 1 + \eta \,,\q\q \eta=const\,.
\ee 
This gives infinitesimal variations on the field space
\be
\delta \xx = \xx \left(\f{\epsilon'}{2}+\eta\right) -\epsilon\, \xx'\,, \q
\delta \cB = \cB \left(\f{\epsilon'}{2}-\eta\right) -\epsilon\, \cB'\,, \label{mob_inf_off}
\ee
and on the solution space
\bsub \be
\delta A_0 &= A_0 \left(\xi -\f{\epsilon'(\phi_0)}{2}\right )\,, \\
\delta \phi_0 &= \epsilon(\phi_0)\,,\\
\delta Q_0 &= -Q_0 \left(\xi +\f{\epsilon'(\psi_0)}{2}\right )\,, \\
\delta \psi_0 &= \epsilon(\psi_0)\,.
\ee \label{mob_inf} \esub

We refer to appendix \ref{app:symm_long} for the transformation laws for $B_0$ and $n$. 

On top of the conformal symmetries \eqref{transf_moebius} there is also another, already known, set of transformations that leave invariant the linear Bondi gauge in two-dimensional dilatonic gravity. This is given by the residual diffeomorphisms \eqref{diffeo}, after imposing the linear radial gauge. A three-parameter family vector field generates them \cite{Ruzziconi:2020wrb},
\be
\Xi = \cE(v) \partial_v + \left( X(v) r + H(v)\right)\partial_r \,.\label{gauge_fixed_vect}
\ee
We use capital letters to distinguish these transformations from the previous ones. The action of these diffeomorphisms can be equivalently thought of as acting on the four-dimensional ansatz \eqref{metric}, or on the two-dimensional plane \eqref{metric2}, with $\xx$ as a scalar field.  Under these, the solution space transforms as 
\bsub \be
\Delta A_0 &= X A_0 +\cE \dot A_0\,,\\
\Delta \phi_0 &= -X \phi_0 +\cE \dot \phi_0-H\,,\\
\Delta B_0 &=2 B_0 \dot \cE +\cE \dot B_0\,,\\
\Delta n &= n\dot \cE +\cE \dot n\,.
\ee \label{diffeo_reduced}\esub
In this case, the expressions take a more compact way in terms of $B_0$, $n$, with respect to $Q_0$ and $\psi_0$. We see immediately that for this transformation the mass is conserved, transforming as a scalar $\delta M= \cE \dot M $. This is not the case for the Moebius transformation \eqref{mob_M}. It is possible to show that the only field-independent transformation that belongs to both classes is the constant translation in space, given by the condition $\epsilon(r) =-H(v)=const$. 

\section{Phase space and conserved charges}
\label{sec:sympl}
In order to discuss the structure of the phase space and the charges associated with the symmetries, we will use the covariant phase space formalism. This has been developed to deal with the role of boundary conditions in gauge theories, highlighting the integrability of charges and their relationship with edge modes \cite{crnkovic_covariant_1986,Lee:1990nz,ashtekar_covariant_1990,Wald:1993nt,Iyer:1994ys,Iyer:1995kg,Jacobson:1993vj,Wald:1999wa}. At the same time, it allows us to deal with the definition of a Poisson structure, through a symplectic structure, for theories with gauge symmetries.  For an action $\cS$, functional of the fields $\chi$, the variation is 
\be
\delta \cS = \int_\cM \rm{EoM}\,  \delta \chi + \int_{\partial \cM} \presym \left( \chi, \delta \chi \right)\,,
\ee
where $\presym$ is the presymplectic potential. It vanishes if we hold fixed some boundary condition on the hypersurface $\partial \cM$, to have a well-defined variational principle. In principle, it might also contain a term of co-dimension two, representing the so-called edge, or corner, modes \cite{Freidel:2020ayo,Freidel:2020svx,Freidel:2020xyx,Freidel:2021cjp,Geiller:2019bti,Speranza:2017gxd,Donnelly:2016auv,Donnelly:2020xgu,Carrozza:2021gju,Carrozza:2022xut}. This contributes to the ambiguity of the definition of the symplectic potential, together with a possible total variation, coming from a boundary Lagrangian and changing the polarisation of the phase space.

\begin{figure}
    \centering
    \includegraphics[width=0.4\linewidth]{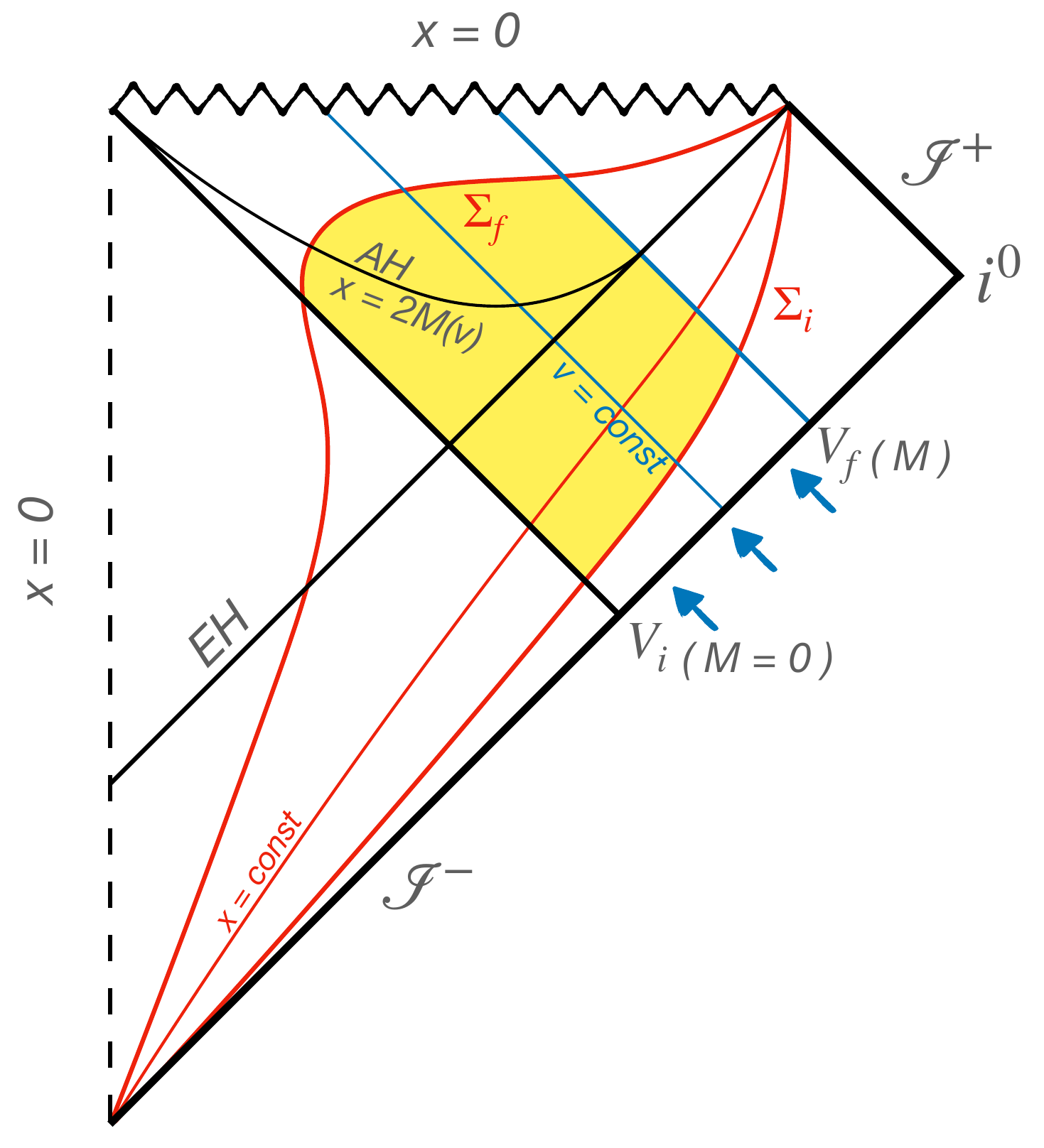}
    \caption{Schematic representation of the Penrose diagram of Vaidya spacetime. We allow here for simplicity $M$ to vary only in the region between $\vi$ and $\vf$, starting from the Minkowski vacuum $M(\vi)=0$, and settling down to a static black hole after $\vf$. The lines at constant radius are drawn in red. While we have the apparent horizon at $\xx=2M$, that coincides with the event horizon only in the static patch. The null boundary $\Gamma$ is then represented by the two surfaces at $\vi$ and $\vf$, while the Cauchy slices are at constant $r$. Let us remark that this is just a pictorial representation because in principle the slices at constant $r$ do not coincide with the ones at constant $\xx$, unless $A_0$ and $\phi_0$ are constants. Our model is well defined in the bulk regardless of the signature of the slices $\Sigma$.}
    \label{fig:Vaidya Penrose}
\end{figure}

Let us start with the first order action \eqref{action_0}. We recall that our field theory is defined in a null stripe between $\vi$ and $\vf$ bounded by two surfaces at constant radius $\ri$ and $\rf$, which, in principle, might be either inside or outside the black hole horizon. The gauge choice allows us to solve the equations of motion explicitly in the radial direction, which makes it the natural evolution parameter (in a Hamiltonian sense) for our theory. This makes us identify the slice at $\ri$ as the Cauchy slice on which we shall set the initial conditions for the variational problem, which then evolves through the other slices $\Sigma(r)$ at a constant radius. The null disconnected boundaries at $\vi$ and $\vf$ are then collectively denoted by $\Gamma$ (see Fig. \ref{fig:Vaidya Penrose}). We can decompose the boundary of the support for the field theory as $\partial \cM = \Sigma(\ri) \cup \Sigma(\rf)\cup \Gamma$. 

The presymplectic potential has two different components, coming from the variation along the radial and null coordinates, these are respectively:
\bsub\be
\presym^r &= \f{\delta \bb \xx' + \delta \xx(\bb'+ 2 \partial_v(N\xx))}{2N \lp^2}\,,\\
\presym^v &= \f{(\xx\delta N  + N\delta \xx) \xx'}{N \lp^2} = \f{\xx'}{N \lp^2}\delta(XN)\,.
\ee\esub
The first is integrated over the slices $\Sigma$, while the second one lives on the boundary $\Gamma$. Once going on-shell of the bulk equations of motion, we can explicitly carry the integration over $r$ and project the whole symplectic potential on a slice $\Sigma$
\be
\Presym &:= \int \dd v \left [\presym^r + \int \dd r\, \partial_v \presym^v \right]\notag\\
&\approx \f{1}{2\lp^2}\int \dd v \left [-\f{\delta B_0}{n} +\f{\dot n \delta(A_0^2\phi_0^2)}{n}-\f{ \partial_v (A_0^2\phi_0^2)\delta n}{n}+ \delta Y\right]\,,\\
\text{with}\;\; Y&= r^2\frac{A_0^2 \dot n}{n}+2r A_0 \left(n-\f{A_0 \phi _0 \dot n}{n}\right)-2A_0 n \phi_0-\partial_v \left(A_0^2 \phi _0^2\right)\,.\notag
\ee
We immediately see that we can renormalize the presymplectic potential, to make it independent of the radius, by eliminating the total variation $Y$, which in any case does not play a role for the symplectic structure $\Sym =\delta\Presym$. The latter turns out to be always conserved along the radial direction, without the need to impose extra boundary conditions on $\Gamma$. By eliminating $Y$, we also recover a well-defined variational principle for the boundary condition (on the Cauchy slice) $\delta B_0 =0$, $\delta n =0$,  $\delta (A_0\phi_0) =0$. The renormalized symplectic form is at the end of the day
\be
\Sym_0 &\approx \int_\vi^\vf  \dd v\left [\f{\delta B_0 \cw \delta n}{2 \lp^2 n^2} +\partial_v\left(\f{\delta(A_0^2 \phi_0^2)\cw \delta n}{2 \lp^2 n} \right) \right]\\
&= \left.\f{\delta(A_0^2 \phi_0^2)\cw \delta n}{2 \lp^2 n}\right|_\vi^\vf +\int_\vi^\vf  \dd v\,\delta M \cw \delta n \notag \,,
\ee
with the field space wedge product $\cw$. We recognise the first part to be a corner term, evaluated on two points in the two-dimensional picture or two opposite homogenous celestial spheres in the four-dimensional point of view. To lighten the notation we will drop the $\vi$ and $\vf$ from the formulas. The subscript $_0$ refers to the fact that we have started from the action $\cS_0$, adding boundary a Lagrangian can change the corner term.  We see that, as should be expected, the only bulk degrees of freedom are the mass and its conjugate momentum which is the null coordinate in the Eddington--Finkelstein gauge (see equation \eqref{on_shell_metric}). To make this clearer, let us consider the bulk symplectic potential
\be
\Theta_0:= \int (n \delta M) \dd v = \int  (\delta M) \dd V\,,\q\q \dd V = n\dd v
\ee
and assume an infalling thin shell, corresponding to a step function mass $M(V)=M \Theta_H(V-V_0)$, for some insertion time of the shell $V_0$. Then we can explicitly integrate over the Cauchy slice and get
\be
\Theta_0:= \int \dd V \left[ \Theta_H(V-V_0) \delta M - \delta(V-V_0) M \delta V_0 \right]
= V_\text{f}\, \delta M -\delta(V_0 M)\,,
\ee
assuming $V_\text{i}<V_0<V_\text{f}$. Discarding the total variation $\delta(V_0 M)$, we see that the conjugate variable to the mass is the null coordinate of the boundary $V_\text{f}$.

As already pointed out, the corner term in the symplectic current depends on the boundary Lagrangian that we choose. For example, if we consider the Einstein--Hilbert action \eqref{action_bg} including the boundary terms, we get
\be
\Sym_\text{EH} \approx \f{\delta(A_0^2 \phi_0^2)\cw \delta n}{4 \lp^2 n}+\f{A_0 \phi_0 \,\delta A_0 \cw\delta \phi_0}{2 \lp^2} +\int \dd v\,\delta M \cw \delta n \,.
\ee

Let us assume the general case (that might correspond to different boundary conditions on the Cauchy slice, e.g Dirichlet, Neumann or mixed)
\be
\Sym \approx \kappa_1 \f{\delta(A_0^2 \phi_0^2)\cw \delta n}{4 \lp^2 n}+\kappa_2 \f{A_0 \phi_0 \,\delta A_0 \cw\delta \phi_0}{2 \lp^2} +\int \dd v\,\delta M \cw \delta n:=\omega_\text{c} +\int \dd v\, \omega_0 \label{sympl_curr}\,,
\ee
with the codimension-2 term $\omega_\text{c}$, relevant to discuss the integrability of large diffeomorphisms, and the codimension-1 term $\omega_0$, capturing the bulk physical degrees of freedom. The presence of local degrees of freedom, represented by the bulk term is a consequence of the \textit{bad} gauge fixing provided by the ansatz \eqref{metric}. The missing mass conservation in the equations of motion, hiding some matter contribution, is translated on the phase space as the seeming emergence of local degrees of freedom from the gravitational action alone. 

With the phase space at hand, we can discuss the realisation of the symmetries \eqref{transf_moebius} and \eqref{diffeo_reduced} on it, the corresponding integrability of charges and their algebra. We will begin with the study of the residual diffeomorphisms.

\subsection{Integrability of the residual diffeomorphisms}
Along the lines of \cite{Ruzziconi:2020wrb}, we will discuss the integrability of the residual diffeomorphisms when they act non-trivially on the boundary $\Gamma$, or equivalently on the corner part of the symplectic current. For this, we shall contract the symplectic form $\Omega$ with the residual spacetime diffeomorphisms \eqref{diffeo_reduced}.

From the codimension-1 term, we immediately see that we can hope to make them integrable only on the solutions with constant mass (non-radiative). Indeed we have 
\be
\Delta_\Xi \cdot \omega_0 = \cE \dot M \delta n - \partial_v  (\cE n) \delta M =\slashed{\delta}Q\,. \label{DOmega0}
\ee
As we stressed before, this is a consequence of ignoring the matter degrees of freedom responsible for the collapse. We leave the problem of including them in the analysis for future works. 

Setting $\dot{M}= 0$, we can search for a \textit{change of slicing} on the phase space that makes the charges integrable. This is also called the Pfaff problem \cite{Barnich:2007bf,Adami:2020ugu,Alessio:2020ioh,Compere:2017knf,Grumiller:2019fmp,Ciambelli:2020shy,Ruzziconi:2020wrb} and amounts to finding a field-dependent choice for the diffeomorphisms parameter \eqref{diffeo_reduced}, such that the variation \eqref{DOmega0} is exact. In general, it is expected that for non-radiative phase spaces (without local degrees of freedom passing through the boundary) such a problem has an infinite number of solutions.

For this, let us take\footnote{Imposing this condition will be denoted by the symbol $\hat =$, as it can be thought of as a non-radiative boundary condition on $\Gamma$} $\partial_v M\,\hat =\,0$. We can easily find a field-dependent parameter that makes the bulk piece integrable. Let us take $\cE =\widetilde \cE/n$
\be
\Delta_{\widetilde{\Xi}} \cdot \omega_0 \,\hat{=}\,  - \partial_v  \widetilde \cE  \delta M\,,
\ee
that gives the mass aspect as the charge associated with reparametrization of the null direction. As a side effect we also see that for constant mass solutions, the last expression in the equation above is turned to a corner term. 

Using the redefinition of $\cE$ to study the corner part of the symplectic current, we get 
\be
\Delta_{\widetilde{\Xi}} \cdot \omega_\text{c} =&\, \widetilde \cE\left( \f{\kappa_1}{4n^2\lp^2}\partial_v(A_0^2 \phi_0^2)\delta n -\kappa_2 \f{\partial_v \phi_0^2 \delta A_0^2-\partial_v A_0^2 \delta \phi_0^2}{8n\lp^2}\right)
+\f{\delta (A_0^2 \phi_0^2)}{4\lp^2} \left(\kappa_2 X- \f{\kappa_1 }{n} \dot{\widetilde \cE} \right)\notag \\
&-H\left(\f{\kappa_1}{2\lp^2 n}A_0^2 \phi_0 \delta n -\f{\kappa_2}{4\lp^2}\phi_0\delta A_0^2 \right)\,.
\ee
We can make this a $\delta$-exact form, by taking the field-dependent transformations
\bsub\be
\cE &=\f{\widetilde \cE}{n}\,,\\
X &=  \widetilde X \f{A_0^2}{\phi_0^2} \left(\f{A_0^2 \phi_0}{n^{\kappa_1/\kappa_2}}\right)^\alpha + \widetilde H \f{A_0^2}{\phi_0^2} \left(\f{A_0^2 \phi_0}{n^{\kappa_1/\kappa_2}}\right)^\beta - \f{\widetilde \cE}{n A_0}\dot A_0  + \f{\kappa_1}{\kappa_2 n}\dot {\widetilde \cE} \,,\\
H &=\widetilde H \f{A_0^2}{\phi_0} \left(\f{A_0^2 \phi_0}{n^{\kappa_1/\kappa_2}}\right)^\beta + \widetilde X \f{A_0^2}{\phi_0} \left(\f{A_0^2 \phi_0}{n^{\kappa_1/\kappa_2}}\right)^\alpha + \f{\widetilde \cE}{n A_0}\partial_v(A_0\phi_0) \,,
\ee\esub
for some real numbers $\alpha$, $\beta$. With this choice, on constant mass solutions, we get the codimension-2 charges
\be
\Delta_{\widetilde \Xi} \cdot \Omega \,\hat =\, \delta\left(\f{\kappa_2}{2\lp^2} \f{\widetilde X}{\alpha} \left(\f{A_0^2 \phi_0}{n^{\kappa_1/\kappa_2}}\right)^\alpha + \f{\kappa_2}{2\lp^2} \f{\widetilde H}{\beta} \left(\f{A_0^2 \phi_0}{n^{\kappa_1/\kappa_2}}\right)^\beta  -\widetilde  \cE M \right)\,.
\ee
The value of the charges seems to depend both on the choice of boundary condition (i.e. the $\kappa$'s) and the particular solution of the Pfaff problem that we choose (i.e. $\alpha$ and $\beta$). However, the charge algebra turns out to be independent of these choices and it is the abelian algebra
\be
\{Q[\widetilde \Xi_1],Q[\widetilde \Xi_2]\} = \Delta_{\widetilde \Xi_1} \cdot \Delta_{\widetilde \Xi_2} \cdot \Omega =0\,. \label{abelian_diff}
\ee
This is consistent with the results in \cite{Ruzziconi:2020wrb}, except for the missing central charge in our case. This difference can have its origin in the different gauge fixing choices. We recall that while here we have fixed $\partial_r N=0$, the choice in \cite{Ruzziconi:2020wrb} and the usual literature about 2d gravity is instead $\partial_r X N)=0$. The different gauge fixing can be interpreted as two different reference frames, and thus two different observers \cite{Carrozza:2021gju,Carrozza:2022xut}. Thus, it is not surprising that the algebras are different as different observers are measuring different physical quantities. A more refined analysis taking into account different gauge fixing and the mapping between them as a change of reference frame is needed to further comment on the comparison with previous results. This is beyond the scope of the present work and we postpone such questions to future works. However, we would like to stress the independence of the abelian algebra on the particular choice of boundary conditions. The latter corresponds to different presymplectic boundary potentials, it is known that the numerical value of the charge can depend on the choice of boundary conditions \cite{Odak:2021axr,Odak:2022ndm,Odak:2023pga}, without affecting their algebra. On the quantum level, this is mapped to the choice of different irreducible representations of the same algebra.

    We would like to end this section with a small remark concerning the edge modes literature. An alternative way of making the charges integrable is to add some edge modes fields, living on the boundary $\Gamma$, that can be interpreted as the image of the residual degrees of freedom leaving in the complementary region outside $\Gamma$. We chose here the approach in \cite{Ruzziconi:2020wrb} of \textit{slice changing}, i.e. considering field-dependent diffeomorphisms, avoiding going too deeply into the construction of reference frames or edge modes \cite{Carrozza:2021gju,Carrozza:2022xut,Goeller:2022rsx}. We expect however the two approaches to be related, as we can usually interpret a field-dependent diffeomorphism as a change of reference frame \cite{Carrozza:2021gju,Carrozza:2022xut}.

\subsection{Conformal transformation and alternative action}

For the Moebius transformation \eqref{transf_moebius}, we lack such freedom of redefining the transformation parameter in a field-dependent way. The finite-dimensional group $\SL(2,\R) \times \SO(2)$ does not allow to take the coefficients of $\epsilon$ and $\xi$ in \eqref{mob_inf} to vary along the null direction. For this, we cannot aim for a \textit{change of slicing} to make the charges integrable and unfortunately, contracting the infinitesimal transformations \eqref{mob_inf} into the symplectic current $\omega$, we find non-integrable quantities. We refer to the appendix \ref{app:symm_long} for the full (lengthy) expression of $\slashed \delta Q$, from which the takeaway message is the non-integrability of the transformation \eqref{transf_moebius}.

This, however, is not too surprising. At some heuristic level, we can see that the Moebius transformations \eqref{mob_inf} and the residual diffeomorphisms  \eqref{diffeo_reduced} look very different. The former is more easily described by the pair of initial conditions $Q_0$ and $\psi_0$, whose \textit{mechanical} interpretation is straightforward in terms of initial value and \textit{velocity} of the field $\cB$, while their spacetime interpretation is more vague. Conversely, $B_0$ and $n$ are nicer geometric quantities, related to the mass and the shell insertion, but their expression in terms of dynamical quantities is more involved. 

More rigorously, this contrast between the two sets of transformations is manifest in the non-covariance of the Moebius transformation, meaning that $\delta_{\{\epsilon,\xi\}}$ and $\Delta_\Xi$ do not form a field independent closed Lie algebra. Although the Moebius transformation \eqref{transf_moebius} is a symmetry of the gauge fixed equations of motion $\delta_{\{\epsilon,\xi\}}\text{EOM}\approx0$, it is not a symmetry of the reduced Lagrangian in the Noether sense. This is not something completely uncommon in physics, even for the very simple model of a free particle, we know that, in general, the conformal rescaling of the position is a symmetry of the equation of motion, but it corresponds to a rescaling of the Lagrangian, not to a total derivative, as required by the Noether theorem.

Nonetheless, we can still associate with the Moebius transformation some conserved quantities along the radial direction. For this, let us consider, on the gauge fixed field space, the following functional
\be
\cS_\text{mob} = \int \dd r \dd v \left [\cB' \xx'\right ] \label{mob_action}\,.
\ee
The corresponding Euler--Lagrange equations are still \eqref{eom_X} and \eqref{eom_B}, the same as for the reduced action coming from general relativity. But, in this case, we lose the constraint \eqref{eom_N}, which however can be obtained from the other two. Indeed assuming $\cB'' \approx 0 \approx \xx''$ we trivially get the conservation along the radial direction of the quantity $\cB' \xx'$, that we can then \textit{define} as $\cN(v)^2$, mimicking the last equation of \eqref{eoms_lin}. From the point of view of the new action functional, there is no gauge freedom, or redundancy in the equation of motion. Equivalently, we can see this by the non-covariance of the functional \eqref{mob_action} under the residual diffeomorphisms generated by \eqref{gauge_fixed_vect}.

From the variational principle of the action \eqref{mob_action} we get
\be
\delta \cS_\text{mob} = -\int \dd r \dd v \left[\cB''\delta \xx + \xx'' \delta \cB \right] + \int \dd v \left[\cB'\delta \xx + \xx' \delta \cB \right]\,,
\ee
and thus the symplectic form 
\be
\Omega_\text{mob} &= \int \dd v\left[ \delta \cB' \cw \delta \xx + \delta \xx' \cw \delta \cB\right]\\
&\approx  \int \dd v\left[ \delta Q_0 \cw \delta(\phi_0 A_0) + \delta A_0 \cw (\psi_0 Q_0) \right]\,. \label{sympl_mob}
\ee
Contracting this with the Moebius transformations \eqref{mob_inf} we get the integrable charges 
\be
\delta_{\{\epsilon,\xi\}} \cdot \Omega_\text{mob} &\approx  \int \dd v \,\delta \left[ \xi Q_0 A_0(\psi_0 - \phi_0) -A_0 Q_0\left( \epsilon_0  +\f{\epsilon_1}{2} (\phi_0+\psi_0) +\epsilon_2 \psi_0 \phi_0\right )\right] \label{Q_moebius}\\
&=  \int \dd v \,\delta \left[ \xi Q_0 A_0(\psi_0 - \phi_0) -\f{A_0 Q_0}{2}\left( \epsilon(\phi_0)  +\epsilon (\psi_0) - (\psi_0-\phi_0)
^2\f{\epsilon''}{2}\right )\right]:= \delta Q[\epsilon,\xi]\,, \notag
\ee
for $\epsilon(r):= \epsilon_0 +\epsilon_1 r +\epsilon_2 r^2$. Their algebra reproduces the Lie algebra of infinitesimal transformations
\be
\{Q[\epsilon_1,\xi_1],Q[\epsilon_2,\xi_2]\} = Q[\epsilon_1 \epsilon_2 '-\epsilon_2 \epsilon_1',\,\xi=0]\,.
\ee
Contrary to the covariant description, in this case, we don't have any edge modes or corner charges, the charges living on the codimension-1 Cauchy slice $\Sigma(r)$. The action \eqref{mob_action} looks like a mechanical action, with a kinetic term quadratic in the radial derivatives, and there is no presence of the null coordinate, except for the integration interval. In other words, the redefinition of the fields, introducing $\cB$, formally maps the Vaidya superspace into an infinite set of decoupled mechanical models isomorphic to the static case, labelled by the null coordinate $v$. The side effect of this construction is the loss of manifest covariance, which can be seen either as the non-covariance of the action \eqref{mob_action} under residual diffeomorphisms \eqref{gauge_fixed_vect} or as the impossibility of making the gauge transformations \eqref{diffeo_reduced} integrable on the symplectic structure $\Omega_\text{mob}$.

Regardless of the non-covariance, we can associate conserved quantities with the Moebius symmetry, in the usual Noether sense. For this, we should work with the  presymplectic potential and infinitesimal transformations on the fields $\cB$ and $\xx$ as in \eqref{mob_inf_off}
\be
\delta_{\{\epsilon,\xi\}} \Omega_\text{mob} 
&\approx \int \dd v\, \delta\left[
\eta(\xx' \cB -\cB' \xx) + \epsilon (\xx' \cB') +\epsilon' (\xx \cB)' +\f{\epsilon''}{2} \xx \cB  
\right] \notag\\
&=\int \dd v\, \delta\left[
\eta(\xx' \cB -\cB' \xx) + \f{1}{2} (\epsilon\xx  \cB)'' -\f{3}{2}\epsilon' (\xx \cB)' 
\right]:= \delta Q\,.
\ee
These are conserved along the radial direction and agree with the on-shell charges given in \eqref{Q_moebius}. Moreover, the interpretation as an infinite set of mechanical models for each null cut is consistent with the fact that the current inside the integral is conserved along the radial direction, even without integrating along the whole Cauchy slice. For each point of $\Sigma(r)$ we can define
\be
j:= \eta(\xx' \cB -\cB' \xx) + \f{1}{2} (\epsilon\xx  \cB)'' -\f{3}{2}\epsilon' (\xx \cB)' \,.
\ee
and we have $j'\approx 0$, corresponding to the Noether charge associated with the Moebius symmetry as in the mechanical setup \cite{Geiller:2022baq, BenAchour:2022fif}.
\subsection{\textit{Heisenberg} extension and full \textit{Schr\"odinger} symmetry}
\label{schro_full}
We can extend the Moebius transformation, mimicking the construction in \cite{Geiller:2022baq, BenAchour:2022fif}. The linearity of the solutions for $\cB$ and $\xx$, makes them trivially invariant under the transformations
\bsub \be
    \xx(r,v) \ &\rightarrow \ \widetilde{\xx}(r,v) =  \xx(r,v) + p_2(v) r + q_1(v)\,,\\
    \cB(r,v) \ &\rightarrow \ \widetilde{\cB}(r,v) = \cB(r,v)+ p_1(v) r + q_2(v)\,, 
\ee \label{transf_heisen}
\esub
They correspond to an abelian symmetry of the equations of motion, for which the infinitesimal and finite transformations coincide. The corresponding action on the initial conditions is 
\bsub \be
\delta A_0 &= p_2 \,, \\
\delta \phi_0 &= -\f{q_1 + \phi_0 p_0}{A_0} \,,\\
\delta Q_0 &= -p_1 \,, \\
\delta \psi_0 &= \f{q_2 + \phi_0 p_2}{Q_0}\,.
\ee \label{heisen_inf} \esub
Using this with the symplectic structure $\Omega_\text{mob}$ also gives an infinite tower of conserved quantities
\be
\delta_{\{\epsilon,\xi\}} \cdot \Omega_\text{mob} &\approx  \int \dd v \,\delta \left[ -q_1 \cB' -q_2 \xx' +p_1 (\xx -r\xx')+p_2 (\cB -r\cB') \right]\\
&\approx  \int \dd v \,\delta \left[p_2 \psi_0 Q_0-p_1 A_0 \phi_0 + q_1 Q_0-q_2 A_0 \right]:= \delta Q[p_i,q_i] \notag\,,
\ee
providing a centrally extended algebra
\be
\{Q[p_i^{(1)},q_i^{(1)}],Q[p_i^{(2)},q_i^{(2)}]\} = \int p_i^{(I)} q_i^{(J)}\epsilon_{IJ}\, \dd v\,,
\ee
with the totally antisymmetric symbol $\epsilon_{IJ}$. As for the Moebius symmetry, we can define here some conserved current on each slice at constant $v$. This means that the centrally extended part in the Schr\"odinger algebra, which is finite-dimensional, is promoted to an infinite dimensional set in the Vaidya model. This can be understood from the mechanical point of view because we can interpret the action \eqref{mob_action}as an infinite set of decoupled mechanical models, one at each null cut $v=const$. We can arbitrarily deform the initial conditions on the slice $\Sigma$ at any point $v$, and the charges $Q[p_i,q_i]$, measures precisely these initial conditions, namely the initial value of the fields $\cB$, $\xx$ and their velocity.
The full algebra is thus
\be
 \left (\sl(2,\R) \oplus \so(2) \right ) \loplus \left (C^\infty(\R^2) \oplus_c C^\infty(\R^2)\right )\,.
\ee
As in the static case, this is an overcomplete set of charges on the phase space. Although the initial conditions are an infinite set of numbers, they can be represented as four (continuous) functions of the null direction. So the charges $Q[p_i,q_i]$ are sufficient to specify the initial value problem and integrate the motion along the radial direction, by exponentiating the charge corresponding to $\epsilon(r)=1$, generator of constant radial translations. It is indeed possible to show that, as in the static setup, here we can obtain the Moebius charges from quadratic combinations of the linear charges $Q[p_i,q_i]$, for example, we have 
\be
Q[\epsilon=1]= Q[q_1=q(v),q_2=0, p_i=0]\, Q[q_1=0,q_2=1/q(v), p_i=0]\,,
\ee
and similar for $\epsilon=r$, $\epsilon=r^2$. 

We would like to stress that all along these sections, we have completely neglected the boundary conditions at the null boundary $\Gamma$. However, restricting to some specific conditions on $\Gamma$ can only affect the fall-off conditions of the initial conditions. Classically, because of the freedom of the dynamics in the null direction, in the bulk, we are always free to set $\{A_0, Q_0,\psi_0, \phi_0\}$ as we desire. The only way of constraining the bulk dynamics is to specify some profile for the infalling null dust, that goes beyond the scope of this work. However, we should remark that once we want to quantize the theory, the fall-off conditions become very relevant, as they affect the structure of the Hilbert space, and consequently the spectrum of some operators. We can imagine that different representations of the symmetry group might correspond to different choices of boundary conditions on $\Gamma$, exploring this direction can give an interesting further development of this work. 
\section*{Discussion}
\label{sec:discussion}
In this article, we have discussed the symmetries of the Vaidya superspace. We have shown that the evolution of a general Vaidya spacetime in the radial direction can be obtained from two different action principles. The first one we have discussed \eqref{action_0} possesses manifestly some residual gauge invariance, leading to codimension-two charges on the corner of the Cauchy slice $\Sigma$. The phase space for this action contains just the mass and the insertion time for a null shell as conjugate variables. All the other degrees of freedom are relevant only at the boundary, playing the role of \textit{edge modes} fields. We have shown that in order to make the corresponding charges integrable, we need to consider field-dependent gauge transformation. Such a choice is not unique, but it always leads to the same abelian charge algebra \eqref{abelian_diff}.

On top of the residual gauge freedom, we have shown that the partial gauge fixed equations of motion \eqref{eoms_lin2} are invariant under a conformal reparametrization of the radius \eqref{transf_moebius}, and an infinite set of linear transformations \eqref{transf_heisen}. These correspond to non-integrable charges on the gravitational reduced phase space, due to the intrinsic non-covariance nature of the transformations. However, we have shown the existence of an alternative action functional \eqref{mob_action} leading to a different phase space, where the extended Schr\"odinger symmetries are integrable. The price to pay is unfortunately the loss of covariance of the theory, which now looks like an infinite set of decoupled mechanical models.

Although both sets of transformations map solutions of the equations of motion in the linear gauge into solutions, they are associated with two different phase spaces. The difference between the two sets is manifest also in the fact that their infinitesimal version separately provides Lie algebras, but together they do not close into a bigger algebra. The commutator of a diffeomorphism and a Moebius transformation gives a new transformation, or equivalently we can say that the structure constants of the two algebras together are field-dependent.

The difference in the phase space is almost irrelevant on the classical level, because the classical equations of motion are the same for the two actions $\cS_\text{EH}$ and $\cS_\text{mob}$, but it becomes crucial once we quantize the theory. A measurement involving quantum processes could then in principle distinguish between the two models.

For the \textit{Moebius action}, the physical degrees of freedom are more numerous than in the general relativity phase space, which is limited to the mass and its conjugated time. The rest of the initial conditions get a physical meaning only on the boundary, as edge modes for the gauge transformation. Conversely in the mechanical setup, they are all already physical in the bulk.

We would like to remark that this discussion does not represent a no-go statement for a generalisation of these conformal symmetries in the full theory, but points out a crucial difference with the boundary large diffeomorphisms, on the contrary to what has been conjectured in previous works \cite{Geiller:2020xze,Sartini:2022ecp}. A more refined analysis including reference frames and (matter) observers could shed light on the origin of the two actions and the physical process underlying this difference.

We stress again that, despite not being integrable on the covariant phase space, the Schr\"odinger symmetries are truly symmetries of the dynamics of Vaidya superspace, mapping solution into solutions and being associated with conserved quantities along the radial direction. As such, they play a role in perturbation theory around the Vaidya background. In the static framework, the conformal reparametrization of the radius is related to the vanishing of Love numbers \cite{BenAchour:2022uqo}, i.e. to the response of black holes to tidal perturbations. The presence of such symmetry in the Vaidya model opens interesting perspectives on the study of its perturbations and the radiative/absorption processes of black holes.

\begin{acknowledgments}
FS is grateful for the hospitality of the association \textit{Esprit des Lieux} in Saint-L\'eger du Ventoux, France, where the initial stages of this work were carried out. The authors would like to thank Marc Geiller, Alejandro Perez and Jibril Ben Achour for the precious discussions and comments. 

This project was made possible through the support of the ID\# 62312 grant from the John Templeton Foundation, as part of the \href{https://www.templeton.org/grant/the-quantum-information-structure-of-spacetime-qiss-second-phase}{\textit{`The Quantum Information Structure of Spacetime'} Project (QISS)}. The opinions expressed in this project/publication are those of the author(s) and do not necessarily reflect the views of the John Templeton Foundation.
\end{acknowledgments}

\newpage
\appendix
\section{Action of the symmetry on the solution space}
\label{app:symm_long}
Let us consider the symmetry transformations as defined in the main text in \eqref{transf_moebius}. These map solutions of the gauge fixed equations \eqref{eom_XX} and \eqref{eom_BB} into solutions. We also recall that the on-shell expressions for the corresponding dynamical fields are of course linear,
\be
\xx\approx A_0(r-\phi_0)\,,\q \cB\approx -Q_0(r-\psi_0)\,.
\ee
Combining this with the symmetry transformation we get the on-shell version of the transformed field, let's take \textit{e.g.} $\xx$,
\be
\widetilde \xx \approx \lambda A_0 \f{r-\phi_0}{\gamma\, r +\delta} = \f{\lambda A_0}{\gamma \phi_0 +\delta}\left(\f{\alpha\, r +\beta}{\gamma\, r+\delta}-\f{\alpha\, \phi_0 +\beta}{\gamma\, \phi_0+\delta}\right) = \lambda\f{A_0}{\sqrt{h'(\phi_0)}} (h(r) -h(\phi_0))  \,,
\ee
from which we easily read the transformation law for $A_0$ and $\phi_0$ presented in the main text's equation \eqref{on_shell_mob}. The exact same calculation for $\cB$ gives us the transformation law for the other two initial conditions $Q_0$ and $\psi_0$.

Concerning the off-shell transformation law for $N$, this must be such that the constraint \eqref{eom_N} is preserved. As shown in the main text, working with the rescaled field $\cN$ is easier. Its value on-shell is given by
\be
\cN^2 \approx -\xx'\cB' \approx n A_0^2(n+ 2\phi_0 \dot A_0) =Q_0 A_0\,,\label{on_shell_cN}
\ee
Now, the constraint \eqref{eom_N} transform as in \eqref{new_constraint}
\be
0& \approx \widetilde \cN^2 +  \partial_{\tilde r} \widetilde\xx \partial_{\tilde r} \widetilde \cB \\
&= \widetilde \cN^2 +\f{1}{h'}\left(\xx' \cB' +\f{1}{2} \partial_r\left(\f{h''}{h'}\xx \cB \right)\right)\notag\\
&\approx \widetilde \cN^2 - \f{A_0\left(B_0 h''(\phi_0)+2 n^2 A_0 h'(\phi_0) +2 n\phi_0 \partial_v (A_0^2 h'(\phi_0))\right)}{2h'(\phi_0)^2}\notag\\
&= \widetilde \cN^2 - \f{Q_0 A_0}{\sqrt{h'(\psi_0)h'(\phi_0)}}=
\widetilde \cN^2 - \widetilde  Q_0 \widetilde A_0\notag\,,
\ee
where in the last line we see that the transformations are indeed symmetries of the constraint \eqref{on_shell_cN}. \newline

Inverting the definition of $Q_0$ and $\psi_0$, and using the variations \eqref{mob_inf}, we can also  write the infinitesimal transformation on the alternative version of the initial conditions $B_0$ and $n$:
\bsub \be
\delta B_0 &= \frac{B_0\left(\phi _0\left(\dot A_0 \left(\epsilon_0'-2 \xi \right)+A_0 \epsilon_0''\dot\phi _0\right)+n \left(\epsilon_0'-2 \xi \right)\right)+2 A_0^2 n^2 \dot \phi _0 \left(\phi _0 \left(\epsilon_0'-4 \xi \right)-2 \epsilon_0\right)}{2 (\phi _0 \dot A_0+n)}\,,\\
\delta n &= -\f{B_0 \epsilon_0''+2 n A_0\left(2 \xi n +\dot A_0\left(2 \epsilon_0+\phi_0(6 \xi - \epsilon_0')\right)\right)}{4 A_0 (n+\phi_0\dot A_0)}\,,
\ee \label{mob_inf_2}\esub 
with the short hand notation $\epsilon_0 = \epsilon(\phi_0)$. Finally, let us remark that this is not a spacetime diffeomorphism, as the mass content of the spacetime is changed, i.e. it does not transform as a scalar field under reparametrization of the null coordinates. We explicitly have
\be
\delta M =& \frac{A_0 \left(\phi_0 \left(A_0\dot \phi_0\left(A_0 \left(\epsilon_0'-4 \xi \right)+\lp^2 M \epsilon_0''\right)+\lp^2 M \dot A_0 \left(10 \xi -\epsilon_0'\right)\right)+\epsilon_0\left(4 \lp^2 M \dot A_0-2 A_0^2 \dot \phi_0\right)\right)}{2 \lp^2 A_0 \left(\phi_0 \dot A_0+n)\right)}\notag\\&+\f{ M n \left(A_0 \left(\epsilon_0'+2 \xi \right)+2 \lp^2 M \epsilon_0''\right)}{2 A_0 \left(\phi_0 \dot A_0+n)\right)} \label{mob_M} .
\ee 
\ \\

Let us finally give the explicit expression for the non-integrability of the Moebius transformation on the symplectic form $\Omega$, separating the codimension-1 and codimension-2 terms we have 
\bsub\be
\delta_{\{\epsilon,\xi\}} \cdot \omega_0 =& \frac{4 A_0 n^2 \left(A_0{}^2 \delta n \dot \phi_0 \left(\phi_0 \left(\epsilon_0'-4 \xi \right)-2 \epsilon_0\right)+\xi  \delta B_0\right)+2 A_0 n \left(\dot A_0 \delta B_0 \left(\phi_0 \left(6 \xi -\epsilon_0'\right)+2 \epsilon_0\right)B_0 \delta n \left(\epsilon_0'-2 \xi \right)\right)}{8 \lp^2 A_0 n^2 \left(\dot A_0 \phi_0+n\right)}\notag\\ 
&+ \f{B_0 \left(\epsilon_0'' \left(2 A_0{}^2 \delta n \phi_0 \dot \phi_0+\delta B_0\right)+2 A_0 \dot A_0 \delta n \phi_0 \left(\epsilon_0'-2 \xi \right)\right)}{8 \lp^2 A_0 n^2 \left(\dot A_0 \phi_0+n\right)},\\
\delta_{\{\epsilon,\xi\}} \cdot \omega_\text{c} =& \f{\phi_0\kappa_1}{8 \lp^2 n \left(\dot A_0 \phi_0+n\right)}  \left(B_0 \delta (A_0 \phi_0) \epsilon_0''+2 \dot A_0 A_0{}^2 \delta n\, \phi_0 \left(\phi_0 \left(2 \xi -\epsilon_0'\right)+2 \epsilon_0\right)\right)\notag  \\
\notag &+ \f{A_0 n\phi_0}{4 \lp^2 \left(\dot A_0 \phi_0+n\right)}\left(A_0 \delta \phi_0 \left(2 \xi  (\kappa_1+\kappa_2)-\kappa_2 \epsilon_0'\right)-2 \kappa_2 \delta A_0 \epsilon_0+2 \kappa_1 \xi  \delta A_0 \phi_0\right) \\
\notag &+ \f{2 A_0 \phi_0^2}{4 \lp^2 \left(\dot A_0 \phi_0+n\right)}\left(\kappa_1 \delta A_0 \dot A_0 \phi_0 \left(6 \xi -\epsilon_0'\right)+A_0 \left(\dot A_0 \delta \phi_0 \left(2 \xi  (3 \kappa_1+\kappa_2)-(\kappa_1+\kappa_2) \epsilon_0'\right)+\kappa_1 \delta n \left(2 \xi -\epsilon_0'\right)\right)\right) \\
 &+  \f{\epsilon_0 A_0 \phi_0}{2 \lp^2  \left(\dot A_0 \phi_0+n\right)}\left( (\kappa_1-\kappa_2) \delta A_0 \dot A_0 \phi_0+\kappa_1 A_0 \left(\dot A_0 \delta \phi_0+\delta n\right)\right).
\ee\esub

\section{Spherically symmetric reduction and useful formulas}
\label{app:2d_dilaton}

\noindent Let us write the spherically symmetric metric as
\be\label{4dSS ansatz}
\dd s^2=g^{(4)}_{\mu\nu}\dd x^\mu\dd x^\nu=g_{ab}\dd x^a\dd x^b+\f{\Phi(t,r)^2}{\lambda^2}\dd\Omega^2,
\ee
where $g_{ab}$ is the 2-metric in the $(t,r)$ plane and $[\lambda]=\text{length}^{-1}$. With this metric we get
\be
\sqrt{-g^{(4)}}=\sqrt{-g}\,\f{\Phi^2}{\lambda^2}\sin\theta,
\q\q
R^{(4)}=R+2\lambda^2\Phi^{-2}-2\Phi^{-2}(\nabla\Phi)^2-4\Phi^{-1}\nabla^2\Phi,
\ee
and therefore the Einstein--Hilbert action becomes
\be\label{4dSS}
S
&=\f{1}{\kappa}\int_M\dd^4x\,\sqrt{-g^{(4)}}\,R^{(4)}\cr
&=\f{1}{\kappa\lambda^2}\int\dd^2x\int_0^\pi\dd\theta\,\sin\theta\int_0^{2\pi}\dd\varphi\,\sqrt{-g}\,\Phi^2R^{(4)}\cr
&=\f{4\pi}{\kappa\lambda^2}\int\dd^2x\,\sqrt{-g}\,\Phi^2\Big(R+2\lambda^2\Phi^{-2}-2\Phi^{-2}(\nabla\Phi)^2-4\Phi^{-1}\nabla^2\Phi\Big)\cr
&=\f{4\pi}{\kappa\lambda^2}\int\dd^2x\,\sqrt{-g}\,\Big(\Phi^2R+2\lambda^2-2(\nabla\Phi)^2-4\Phi\nabla^2\Phi\Big)\cr
&=\f{4\pi}{\kappa\lambda^2}\int\dd^2x\,\sqrt{-g}\,\Big(\Phi^2R+2\lambda^2+2(\nabla\Phi)^2-4\nabla_a(\Phi\nabla^a\Phi)\Big),
\ee
where later on we will set $\kappa=16\pi$. The action for a minimally coupled massless scalar field $f$ is
\be
S_\text{m}=-\int_M\dd^4x\,\sqrt{-g^{(4)}}\,(\nabla f)^2=-\f{2\pi}{\lambda^2}\int\dd^2x\,\sqrt{-g}\,\Phi^2(\nabla f)^2.
\ee

Under variations we have
\bsub
\be
\delta\Gamma^\rho_{\mu\nu}&=\f{1}{2}g^{\rho\sigma}(\nabla_\mu\delta g_{\sigma\nu}+\nabla_\nu\delta g_{\sigma\mu}-\nabla_\sigma\delta g_{\mu\nu}),\\
\delta R_{\mu\nu}&=\nabla_\rho\delta\Gamma^\rho_{\mu\nu}-\nabla_\nu\delta\Gamma^\rho_{\mu\rho}=\f{1}{2}(\nabla^\rho\nabla_\mu\delta g_{\rho\nu}+\nabla^\rho\nabla_\nu\delta g_{\mu\rho}-g^{\rho\sigma}\nabla_\mu\nabla_\nu\delta g_{\rho\sigma}-\nabla^2\delta g_{\mu\nu}),\\
\delta R
&=\delta g^{\mu\nu}R_{\mu\nu}+g^{\mu\nu}\delta R_{\mu\nu}\cr
&=\delta g^{\mu\nu}R_{\mu\nu}+\nabla_\mu(g^{\rho\sigma}\delta\Gamma^\mu_{\rho\sigma}-g^{\rho\mu}\delta\Gamma^\nu_{\rho\nu})\cr
&=\delta g^{\mu\nu}R_{\mu\nu}+\nabla^\mu\nabla^\nu(\delta g_{\mu\nu}-g_{\mu\nu}g^{\rho\sigma}\delta g_{\rho\sigma})\cr
&=\delta g^{\mu\nu}R_{\mu\nu}+(\nabla^\mu\nabla^\nu-g^{\mu\nu}\nabla^2)\delta g_{\mu\nu}\cr
&=\delta g^{\mu\nu}R_{\mu\nu}+\nabla_\mu\nabla_\nu(g^{\mu\nu}g_{\rho\sigma}\delta g^{\rho\sigma}-\delta g^{\mu\nu}),\\
\delta\sqrt{-g}&=\f{1}{2}\sqrt{-g}\,g^{\mu\nu}\delta g_{\mu\nu}.
\ee
\esub
Under a conformal rescaling $g_{\mu\nu}=\Omega^2\tilde{g}_{\mu\nu}=e^{2\sigma}\tilde{g}_{\mu\nu}$ we have 
\bsub
\be
\sqrt{-g}&=\Omega^2\sqrt{-\tilde{g}},\\
R&=\Omega^{-2}\big(\tilde{R}-2\Omega^{-1}\tilde{\nabla}^2\Omega+2\Omega^{-2}(\tilde{\nabla}\Omega)^2\big)=e^{-2\sigma}\big(\tilde{R}-2\tilde{\nabla}^2\sigma\big),\\
\nabla_\mu\phi&=\tilde{\nabla}_\mu\phi,\\
(\nabla\phi)^2&=\Omega^{-2}(\tilde{\nabla}\phi)^2,\\
\nabla^2\phi&=\Omega^{-2}\tilde{\nabla}^2\phi,\\
\nabla_\mu\nabla_\nu\phi&=\tilde{\nabla}_\mu\tilde{\nabla}_\nu\phi-\tilde{\nabla}_\mu\sigma\tilde{\nabla}_\nu\phi-\tilde{\nabla}_\mu\phi\tilde{\nabla}_\nu\sigma+\tilde{g}_{\mu\nu}\tilde{\nabla}^\alpha\sigma\tilde{\nabla}_\alpha\phi.
\ee
\esub
With this, for a Lagrangian of the form
\be
L=\sqrt{-g}\,\Big(V(\Phi)R+U(\Phi)+W(\Phi)(\nabla\Phi)^2\Big),
\ee
we can always remove the kinetic term for $\Phi$ by using
\be
\Omega(\Phi)=\exp\left(-\int_{\Phi_0}^\Phi\f{W(z)}{2V'(z)}\dd z\right).
\ee

\bibliographystyle{Biblio}
\bibliography{Biblio}

\providecommand{\href}[2]{#2}\begingroup\raggedright\begin{thebibliography}{10}

\bibitem{Regge:1974zd}
T.~Regge and C.~Teitelboim, \emph{Role of {Surface} {Integrals} in the
  {Hamiltonian} {Formulation} of {General} {Relativity}},
  \href{http://dx.doi.org/10.1016/0003-4916(74)90404-7}{\emph{Annals Phys.}
  {\bfseries 88} (1974) 286}.

\bibitem{Carlip:1994gy}
S.~Carlip, \emph{The {Statistical} {Mechanics} of the (2+1)-{Dimensional}
  {Black} {Hole}}, \href{http://dx.doi.org/10.1103/PhysRevD.51.632}{\emph{Phys.
  Rev. D} {\bfseries 51} (1995) 632--637}.

\bibitem{Balachandran:1995qa}
A.~P. Balachandran, L.~Chandar and A.~Momen, \emph{Edge {States} in {Canonical}
  {Gravity}}, .

\bibitem{Freidel:2020xyx}
L.~Freidel, M.~Geiller and D.~Pranzetti, \emph{Edge modes of gravity -- {I}:
  {Corner} potentials and charges},
  \href{http://dx.doi.org/10.1007/JHEP11(2020)026}{\emph{JHEP} {\bfseries 11}
  (Nov., 2020) 026}.

\bibitem{Freidel:2020svx}
L.~Freidel, M.~Geiller and D.~Pranzetti, \emph{Edge modes of gravity -- {II}:
  {Corner} metric and {Lorentz} charges},
  \href{http://dx.doi.org/10.1007/JHEP11(2020)027}{\emph{JHEP} {\bfseries 11}
  (Nov., 2020) 027}.

\bibitem{Freidel:2020ayo}
L.~Freidel, M.~Geiller and D.~Pranzetti, \emph{Edge modes of gravity. {Part}
  {III}. {Corner} simplicity constraints},
  \href{http://dx.doi.org/10.1007/JHEP01(2021)100}{\emph{JHEP} {\bfseries 01}
  (Jan., 2021) 100}.

\bibitem{Donnelly:2020xgu}
W.~Donnelly, L.~Freidel, S.~F. Moosavian and A.~J. Speranza,
  \emph{Gravitational edge modes, coadjoint orbits, and hydrodynamics},
  \href{http://dx.doi.org/10.1007/JHEP09(2021)008}{\emph{JHEP} {\bfseries 09}
  (Sept., 2021) 008}.

\bibitem{Freidel:2021cjp}
L.~Freidel, R.~Oliveri, D.~Pranzetti and S.~Speziale, \emph{Extended corner
  symmetry, charge bracket and {Einstein}'s equations},
  \href{http://dx.doi.org/10.1007/JHEP09(2021)083}{\emph{JHEP} {\bfseries 09}
  (Sept., 2021) 083}.

\bibitem{Carrozza:2021gju}
S.~Carrozza and P.~A. Hoehn, \emph{Edge modes as reference frames and boundary
  actions from post-selection},
  \href{http://dx.doi.org/10.1007/JHEP02(2022)172}{\emph{JHEP} {\bfseries 02}
  (Feb., 2022) 172}.

\bibitem{Carrozza:2022xut}
S.~Carrozza, S.~Eccles and P.~A. Hoehn, \emph{Edge modes as dynamical frames:
  charges from post-selection in generally covariant theories}, .

\bibitem{Newman:1965tw}
E.~T. Newman and A.~I. Janis, \emph{Note on the {Kerr} {Spinning}-{Particle}
  {Metric}}, \href{http://dx.doi.org/10.1063/1.1704350}{\emph{J. Math. Phys.}
  {\bfseries 6} (1965) 915--917}.

\bibitem{Drake:1998gf}
S.~P. D.~P. Szekeres, \emph{An explanation of the {Newman}-{Janis}
  {Algorithm}}, \href{http://dx.doi.org/10.1023/A:1001920232180}{\emph{Gen.
  Rel. Grav.} {\bfseries 32} (2000) 445--458}.

\bibitem{Chen:2010ik}
B.~Chen and J.~Long, \emph{Hidden {Conformal} {Symmetry} and {Quasi}-normal
  {Modes}}, \href{http://dx.doi.org/10.1103/PhysRevD.82.126013}{\emph{Phys.
  Rev. D} {\bfseries 82} (2010) 126013}.

\bibitem{Kim:2012mh}
Y.-W. Kim, Y.~S. Myung and Y.-J. Park, \emph{Quasinormal modes and hidden
  conformal symmetry in the {Reissner}-{Nordstrom} black hole},
  \href{http://dx.doi.org/10.1140/epjc/s10052-013-2440-8}{\emph{Eur. Phys. J.
  C} {\bfseries 73} (2013) 2440}.

\bibitem{Birmingham:1998jt}
D.~Birmingham, I.~Sachs and S.~Sen, \emph{Entropy of {Three}-{Dimensional}
  {Black} {Holes} in {String} {Theory}},
  \href{http://dx.doi.org/10.1016/S0370-2693(98)00236-6}{\emph{Phys. Lett. B}
  {\bfseries 424} (1998) 275--280}.

\bibitem{Carlip:1998wz}
S.~Carlip, \emph{Black {Hole} {Entropy} from {Conformal} {Field} {Theory} in
  {Any} {Dimension}},
  \href{http://dx.doi.org/10.1103/PhysRevLett.82.2828}{\emph{Phys. Rev. Lett.}
  {\bfseries 82} (1999) 2828--2831}.

\bibitem{Carlip:2017xne}
S.~Carlip, \emph{Black {Hole} {Entropy} from {BMS} {Symmetry} at the
  {Horizon}},
  \href{http://dx.doi.org/10.1103/PhysRevLett.120.101301}{\emph{Phys. Rev.
  Lett.} {\bfseries 120} (Mar., 2018) 101301}.

\bibitem{Geiller:2020xze}
M.~Geiller, E.~R. Livine and F.~Sartini, \emph{Symmetries of the black hole
  interior and singularity regularization},
  \href{http://dx.doi.org/10.21468/SciPostPhys.10.1.022}{\emph{SciPost Phys.}
  {\bfseries 10} (Jan., 2021) 022}.

\bibitem{BenAchour:2022fif}
J.~Ben~Achour, E.~R. Livine, D.~Oriti and G.~Piani, \emph{Schr{\"o}dinger
  {Symmetry} in {Gravitational} {Mini}-{Superspaces}}, .

\bibitem{Geiller:2022baq}
M.~Geiller, E.~R. Livine and F.~Sartini, \emph{Dynamical symmetries of
  homogeneous minisuperspace models},
  \href{http://dx.doi.org/10.1103/PhysRevD.106.064013}{\emph{Phys. Rev. D}
  {\bfseries 106} (Sept., 2022) 064013}.

\bibitem{BenAchour:2023dgj}
J.~Ben~Achour, E.~R. Livine and D.~Oriti, \emph{Schr{\"o}dinger symmetry of
  {Schwarzschild}-({A}){dS} black hole mechanics}, .

\bibitem{BenAchour:2017qpb}
J.~Ben~Achour and E.~R. Livine, \emph{Thiemann complexifier in classical and
  quantum {FLRW} cosmology},
  \href{http://dx.doi.org/10.1103/PhysRevD.96.066025}{\emph{Phys. Rev. D}
  {\bfseries 96} (Sept., 2017) 066025}.

\bibitem{BenAchour:2019ywl}
J.~Ben~Achour and E.~R. Livine, \emph{Protected {SL}(2,{R}) {Symmetry} in
  {Quantum} {Cosmology}},
  \href{http://dx.doi.org/10.1088/1475-7516/2019/09/012}{\emph{JCAP} {\bfseries
  09} (Sept., 2019) 012}.

\bibitem{Sartini:2022ecp}
F.~Sartini, \emph{Hidden {Symmetries} in {Gravity} : {Black} holes and other
  minisuperspaces}.
\newblock {PhD} {Thesis}, Laboratoire de Physique de l'ENS Lyon, France, ENS,
  Lyon, Lab. Phys., July, 2022.

\bibitem{Cariglia:2016oft}
M.~Cariglia, C.~Duval, G.~W. Gibbons and P.~A. Horvathy, \emph{Eisenhart lifts
  and symmetries of time-dependent systems},
  \href{http://dx.doi.org/10.1016/j.aop.2016.07.033}{\emph{Annals Phys.}
  {\bfseries 373} (Oct., 2016) 631--654}.

\bibitem{Bicak:1997bx}
J.~Bicak and K.~Kuchar, \emph{Null dust in canonical gravity},
  \href{http://dx.doi.org/10.1103/PhysRevD.56.4878}{\emph{Phys. Rev. D}
  {\bfseries 56} (1997) 4878--4895}.

\bibitem{Louko:1997wc}
J.~Louko, B.~F. Whiting and J.~L. Friedman, \emph{Hamiltonian spacetime
  dynamics with a spherical null-dust shell},
  \href{http://dx.doi.org/10.1103/PhysRevD.57.2279}{\emph{Phys. Rev. D}
  {\bfseries 57} (1998) 2279--2298}.

\bibitem{Campiglia:2016fzp}
M.~Campiglia, R.~Gambini, J.~Olmedo and J.~Pullin, \emph{Quantum
  self-gravitating collapsing matter in a quantum geometry},
  \href{http://dx.doi.org/10.1088/0264-9381/33/18/18LT01}{\emph{Class. Quant.
  Grav.} {\bfseries 33} (Aug., 2016) 18LT01}.

\bibitem{Hajicek:2002ny}
P.~Hajicek, \emph{Quantum theory of gravitational collapse (lecture notes on
  quantum conchology)},  in \emph{Lect. {Notes} {Phys}.}, vol.~631,
  pp.~255--299.
\newblock 2003.
\newblock \href{http://dx.doi.org/10.1007/978-3-540-45230-0_6}{DOI}.

\bibitem{Eyheralde:2017jzd}
R.~Eyheralde, M.~Campiglia, R.~Gambini and J.~Pullin, \emph{Quantum fluctuating
  geometries and the information paradox},
  \href{http://dx.doi.org/10.1088/1361-6382/aa8e30}{\emph{Class. Quant. Grav.}
  {\bfseries 34} (Nov., 2017) 235015}.

\bibitem{Grumiller:2002nm}
D.~Grumiller, W.~Kummer and D.~V. Vassilevich, \emph{Dilaton {Gravity} in {Two}
  {Dimensions}},
  \href{http://dx.doi.org/10.1016/S0370-1573(02)00267-3}{\emph{Phys. Rept.}
  {\bfseries 369} (2002) 327--430}.

\bibitem{Afshar:2019axx}
H.~Afshar, H.~A. Gonz{\'a}lez, D.~Grumiller and D.~Vassilevich, \emph{Flat
  space holography and the complex {Sachdev}-{Ye}-{Kitaev} model},
  \href{http://dx.doi.org/10.1103/PhysRevD.101.086024}{\emph{Phys. Rev. D}
  {\bfseries 101} (Apr., 2020) 086024}.

\bibitem{Ruzziconi:2020wrb}
R.~Ruzziconi and C.~Zwikel, \emph{Conservation and {Integrability} in
  {Lower}-{Dimensional} {Gravity}},
  \href{http://dx.doi.org/10.1007/JHEP04(2021)034}{\emph{JHEP} {\bfseries 04}
  (Apr., 2021) 034}.

\bibitem{Grumiller:2021cwg}
D.~Grumiller, R.~Ruzziconi and C.~Zwikel, \emph{Generalized dilaton gravity in
  2d}, \href{http://dx.doi.org/10.21468/SciPostPhys.12.1.032}{\emph{SciPost
  Phys.} {\bfseries 12} (Jan., 2022) 032}.

\bibitem{crnkovic_covariant_1986}
C.~Crnkovic and E.~Witten, \emph{{COVARIANT} {DESCRIPTION} {OF} {CANONICAL}
  {FORMALISM} {IN} {GEOMETRICAL} {THEORIES}}, .

\bibitem{Lee:1990nz}
J.~Lee and R.~M. Wald, \emph{Local symmetries and constraints},
  \href{http://dx.doi.org/10.1063/1.528801}{\emph{J. Math. Phys.} {\bfseries
  31} (1990) 725--743}.

\bibitem{ashtekar_covariant_1990}
A.~Ashtekar, L.~Bombelli and O.~Reula, \emph{{THE} {COVARIANT} {PHASE} {SPACE}
  {OF} {ASYMPTOTICALLY} {FLAT} {GRAVITATIONAL} {FIELDS}}, .

\bibitem{Wald:1993nt}
R.~M. Wald, \emph{Black {Hole} {Entropy} is {Noether} {Charge}},
  \href{http://dx.doi.org/10.1103/PhysRevD.48.R3427}{\emph{Phys. Rev. D}
  {\bfseries 48} (Oct., 1993) R3427--R3431}.

\bibitem{Iyer:1994ys}
V.~Iyer and R.~M. Wald, \emph{Some {Properties} of {Noether} {Charge} and a
  {Proposal} for {Dynamical} {Black} {Hole} {Entropy}},
  \href{http://dx.doi.org/10.1103/PhysRevD.50.846}{\emph{Phys. Rev. D}
  {\bfseries 50} (1994) 846--864}.

\bibitem{Iyer:1995kg}
V.~Iyer and R.~M. Wald, \emph{A comparison of {Noether} charge and {Euclidean}
  methods for {Computing} the {Entropy} of {Stationary} {Black} {Holes}},
  \href{http://dx.doi.org/10.1103/PhysRevD.52.4430}{\emph{Phys. Rev. D}
  {\bfseries 52} (1995) 4430--4439}.

\bibitem{Jacobson:1993vj}
T.~Jacobson, G.~Kang and R.~C. Myers, \emph{On {Black} {Hole} {Entropy}},
  \href{http://dx.doi.org/10.1103/PhysRevD.49.6587}{\emph{Phys. Rev. D}
  {\bfseries 49} (1994) 6587--6598}.

\bibitem{Wald:1999wa}
R.~M. Wald and A.~Zoupas, \emph{A {General} {Definition} of "{Conserved}
  {Quantities}" in {General} {Relativity} and {Other} {Theories} of {Gravity}},
  \href{http://dx.doi.org/10.1103/PhysRevD.61.084027}{\emph{Phys. Rev. D}
  {\bfseries 61} (2000) 084027}.

\bibitem{Geiller:2019bti}
M.~Geiller and P.~Jai-akson, \emph{Extended actions, dynamics of edge modes,
  and entanglement entropy},
  \href{http://dx.doi.org/10.1007/JHEP09(2020)134}{\emph{JHEP} {\bfseries 09}
  (Sept., 2020) 134}.

\bibitem{Speranza:2017gxd}
A.~J. Speranza, \emph{Local phase space and edge modes for
  diffeomorphism-invariant theories},
  \href{http://dx.doi.org/10.1007/JHEP02(2018)021}{\emph{JHEP} {\bfseries 02}
  (Feb., 2018) 021}.

\bibitem{Donnelly:2016auv}
W.~Donnelly and L.~Freidel, \emph{Local subsystems in gauge theory and
  gravity}, \href{http://dx.doi.org/10.1007/JHEP09(2016)102}{\emph{JHEP}
  {\bfseries 09} (Sept., 2016) 102}.

\bibitem{Barnich:2007bf}
G.~Barnich and G.~Compere, \emph{Surface charge algebra in gauge theories and
  thermodynamic integrability},
  \href{http://dx.doi.org/10.1063/1.2889721}{\emph{J. Math. Phys.} {\bfseries
  49} (2008) 042901}.

\bibitem{Adami:2020ugu}
H.~Adami, M.~M. Sheikh-Jabbari, V.~Taghiloo, H.~Yavartanoo and C.~Zwikel,
  \emph{Symmetries at {Null} {Boundaries}: {Two} and {Three} {Dimensional}
  {Gravity} {Cases}},
  \href{http://dx.doi.org/10.1007/JHEP10(2020)107}{\emph{JHEP} {\bfseries 10}
  (Oct., 2020) 107}.

\bibitem{Alessio:2020ioh}
F.~Alessio, G.~Barnich, L.~Ciambelli, P.~Mao and R.~Ruzziconi, \emph{Weyl
  {Charges} in {Asymptotically} {Locally} {AdS}\$\_3\$ {Spacetimes}},
  \href{http://dx.doi.org/10.1103/PhysRevD.103.046003}{\emph{Phys. Rev. D}
  {\bfseries 103} (Feb., 2021) 046003}.

\bibitem{Compere:2017knf}
G.~Comp{\`e}re and A.~Fiorucci, \emph{Asymptotically flat spacetimes with
  {BMS}\$\_3\$ symmetry},
  \href{http://dx.doi.org/10.1088/1361-6382/aa8aad}{\emph{Class. Quant. Grav.}
  {\bfseries 34} (Sept., 2017) 204002}.

\bibitem{Grumiller:2019fmp}
D.~Grumiller, A.~P{\'e}rez, M.~M. Sheikh-Jabbari, R.~Troncoso and C.~Zwikel,
  \emph{Spacetime structure near generic horizons and soft hair},
  \href{http://dx.doi.org/10.1103/PhysRevLett.124.041601}{\emph{Phys. Rev.
  Lett.} {\bfseries 124} (Jan., 2020) 041601}.

\bibitem{Ciambelli:2020shy}
L.~Ciambelli, S.~Detournay and A.~Somerhausen, \emph{New {Chiral} {Gravity}},
  \href{http://dx.doi.org/10.1103/PhysRevD.102.106017}{\emph{Phys. Rev. D}
  {\bfseries 102} (Nov., 2020) 106017}.

\bibitem{Odak:2021axr}
G.~Odak and S.~Speziale, \emph{Brown-{York} charges with mixed boundary
  conditions}, \href{http://dx.doi.org/10.1007/JHEP11(2021)224}{\emph{JHEP}
  {\bfseries 11} (Nov., 2021) 224}.

\bibitem{Odak:2022ndm}
G.~Odak, A.~Rignon-Bret and S.~Speziale, \emph{Wald-{Zoupas} prescription with
  (soft) anomalies},
  \href{http://dx.doi.org/10.1103/PhysRevD.107.084028}{\emph{Phys. Rev. D}
  {\bfseries 107} (Apr., 2023) 084028}.

\bibitem{Odak:2023pga}
G.~Odak, A.~Rignon-Bret and S.~Speziale, \emph{General gravitational charges on
  null hypersurfaces}, .

\bibitem{Goeller:2022rsx}
C.~Goeller, P.~A. Hoehn and J.~Kirklin, \emph{Diffeomorphism-invariant
  observables and dynamical frames in gravity: reconciling bulk locality with
  general covariance}, .

\bibitem{BenAchour:2022uqo}
J.~B. Achour, E.~R. Livine, S.~Mukohyama and J.-P. Uzan, \emph{Hidden symmetry
  of the static response of black holes: {Applications} to {Love} numbers}, .

\end{thebibliography}\endgroup

\end{document}